\documentclass[twocolumn]{IEEEtran}

\usepackage{amsmath,graphics,amssymb,graphicx}
\usepackage{subfigure}
\usepackage{url}
\usepackage{color,soul}
\usepackage{verbatim}
\usepackage{amsthm}
\usepackage{tabularx}
\usepackage{cite}
\usepackage{multirow}
\usepackage{subfigure}
\usepackage{booktabs}

\newtheorem{prop}{Proposition}

\begin{document}
\setlength{\abovedisplayskip}{5pt}
\setlength{\belowdisplayskip}{5pt}
\title{{Self-Nomination: Deep Learning for Decentralized CSI Feedback Reduction in MU-MIMO Systems}}

% \author{\IEEEauthorblockN{Juseong~Park \emph{et al.}} \\ 
\author{\IEEEauthorblockN{Juseong~Park, Foad~Sohrabi, Jinfeng~Du, and Jeffrey~G.~Andrews} \\ 
% \vspace*{-6mm}
% \thanks{Copyright (c) 2015 IEEE. Personal use of this material is permitted. However, permission to use this material for any other purposes must be obtained from the IEEE by sending a request to pubs-permissions@ieee.org.}
% \thanks{This work was supported by. \textit{(Corresponding authors: .)}
% \thanks{This work has been supported by Nokia Bell Labs and in part by the National Science Foundation RINGS program grant CNS-2148141.}
\thanks{This work has been supported by Nokia Bell Labs and in part by the National Science Foundation RINGS program grant CNS-2148141.}
\thanks{An early and partial version of this work was presented in \cite{Park24_asilomar}.}
\thanks{Juseong Park and Jeffrey G. Andrews are with 6G@UT in the Wireless Networking and Communications Group, The University of Texas at Austin, Austin, TX 78712, USA (email: juseong.park@utexas.edu; jandrews@ece.utexas.edu).}
\thanks{Foad Sohrabi and Jinfeng Du are with Radio Systems Research, Nokia Bell Labs, Murray Hill, NJ 07974 USA (e-mail: foad.sohrabi@nokia-bell-labs.com; jinfeng.du@nokia-bell-labs.com).}
% \thanks{Last modified: \today.}
}
\maketitle


% \maketitle

\begin{abstract}
This paper introduces a novel deep learning-based user-side feedback reduction framework, termed \emph{self-nomination}. 
The goal of self-nomination is to reduce the number of users (UEs) feeding back channel state information (CSI) to the base station (BS), 
by letting each UE decide whether to feed back based on its estimated likelihood of being scheduled and its potential contribution to precoding in a multiuser MIMO (MU-MIMO) downlink.
% while ensuring the users that do feedback have a high probability of being scheduled in a multiuser MIMO (MU-MIMO) downlink. 
% Unlike thresholding methods that make feedback decisions based on SNR or SINR, 
Unlike SNR- or SINR-based thresholding methods, the proposed approach uses rich spatial channel statistics and learns nontrivial correlation effects that affect eventual MU-MIMO scheduling decisions.
% To train the self-nomination neural network under an average sum feedback rate constraint, we propose two optimization strategies. The first employs direct gradient descent on a Lagrangian reformulation of the sum-rate objective, leveraging a primal-dual method with gradient approximations. The second circumvents these gradient approximations through policy gradient-based optimization with a stochastic Bernoulli policy, accommodating non-differentiable user scheduling without requiring differentiation. The proposed framework is extended to proportional-fair scheduling by incorporating dynamic per-user weights into the feedback decisions.
To train the self-nomination network under an average feedback constraint, we propose two different strategies: one based on direct optimization with gradient approximations, and another using policy gradient-based optimization with a stochastic Bernoulli policy to handle non-differentiable scheduling. The framework also supports proportional-fair scheduling by incorporating dynamic user weights.
% Under realistic 3GPP channel models, 
Numerical results confirm that the proposed self-nomination method significantly reduces CSI feedback overhead.
% and the number of UEs performing feedback.  
Compared to baseline feedback methods, self-nomination can reduce feedback by as much as 65\%, saving not only bandwidth but also allowing many UEs to avoid feedback altogether (and thus, potentially enter a sleep mode).  Self-nomination achieves this significant savings with negligible reduction in sum-rate or fairness.  
\end{abstract}


\section{Introduction} \label{sec:introduction}

The key new spectrum for 6G will be so-called Frequency Range 3 (FR3), particularly 7-15 GHz. The wavelengths in this range enable large antenna arrays compared to FR1 (Sub-6GHz), yet more favorable propagation characteristics and larger channel ranks as compared to FR2, the millimeter wave bands \cite{Zhang25, Miao23}.
As a result, massive multiple-input multiple-output (MIMO) technology remains a central component of 6G, with a shift away from beamforming-centric architectures, which only focus on a single strong beam, toward a more native MIMO paradigm that more effectively exploits spatial multiplexing \cite{Andrews25}. At the same time, the rapid growth of fixed wireless access (FWA) has increased interest in multiuser MIMO (MU-MIMO) as a key enabler of sustained high-throughput communication to fixed devices, which are more amenable to MU-MIMO than to mobile devices with sporadic throughput demands. %In order to fully take advantage of MU-MIMO’s capabilities, however, we must solve several real-world problems, especially when the system becomes large.}

Downlink MU-MIMO requires timely and accurate CSI feedback from the UEs, in order to compute high performance precoders.  Even TDD systems exploiting channel reciprocity do not escape this requirement, since the uplink (UL) transmit power is so much lower than in the downlink (precluding accurate wideband BS-side channel estimation), and some aspects of the channel (such as inter-cell interference) do not obey reciprocity.   Furthermore, even the uplink pilots used for reciprocal downlink channel estimation can be viewed as a type of feedback, which consumes bandwidth and power.  As such, current cellular and Wi-Fi standards require user equipment (UE) to provide channel state information (CSI) periodically, regardless of their scheduling likelihood, leading to unnecessary transmissions and power consumption on the UE side, and increased scheduling complexity at the base station (BS). As the number of UEs grows, these inefficiencies scale unfavorably, particularly in MU-MIMO, where user scheduling and interference management further complicate system design \cite{Zukang06, Castaneda17}. %\blue{Unfortunately, leveraging channel reciprocity in time division duplex (TDD) systems cannot serve as a general solution, since cell-edge UEs often lack sufficient power for accurate uplink channel estimation. Moreover, this approach merely shifts the feedback burden from UEs to the BS without reducing the overall signaling overhead.}
% Efforts have been made to improve the efficiency of conventional feedback methods, such as more effective CSI compression, but these approaches remain insufficient in large-scale deployments. As networks continue to expand, innovative solutions are required to reduce feedback overhead while maintaining system performance.
This paper proposes and analyzes a distributed UE-side feedback reduction method that is designed to address these issues. Each UE autonomously decides whether to report CSI or remain silent based on its instantaneous downlink channel conditions and predicted scheduling likelihood. The proposed mechanism aims to reduce uplink 
% control channel transmission, 
transmissions for CSI feedback,
while improving both system efficiency and UE energy consumption.

\subsection{Related Work}
Various methods for feedback decision-making at the UE side have been explored in prior work. 
In an early study \cite{Gesbert04}, the authors introduced signal-to-noise ratio (SNR)-based thresholding for multiuser diversity in single-input single-output systems, allowing only users whose SNR exceeded a certain threshold to feed back. This approach demonstrated the potential benefits of selective UE feedback. Later extensions incorporated multiple SNR thresholds \cite{Hassel05, Hassel07} and generalized the approach to MIMO by replacing SNR with signal-to-interference-plus-noise ratio (SINR) under random beamforming \cite{Pugh10, Pugh11, Chang15}. Other studies investigated CSI quantization for multi-threshold MIMO feedback \cite{Li14} and examined feedback reduction techniques for antenna selection in one-bit quantization MIMO systems \cite{Zhang07}. Additionally, sum feedback rate constraints were introduced to reflect uplink capacity for CSI feedback \cite{Huang07}.

While these studies paved the way for selective feedback, many rely on non-standardized techniques, such as random beamforming and antenna selection, that are suboptimal and dissimilar from modern wireless standards. Moreover, many of these methods assume simplified channel models, such as independent and identically distributed (i.i.d.) Rayleigh fading, which are far from real-world deployments. The SINR-based thresholding method, in particular, fails to account for factors like beamforming design and spatial channel characteristics. Furthermore, most existing approaches primarily focus on asymptotic multiuser diversity gains as the number of UEs increases indefinitely, leaving unresolved challenges in scenarios with moderately sized user populations, which are more typical in practice.

Meanwhile, deep learning-based CSI compression methods have been actively investigated to reduce CSI feedback overhead in MIMO systems, forming a class of techniques potentially complementary to selective feedback, and have been formally addressed by the 3rd Generation Partnership Project (3GPP) \cite{3gpp2024, Wang25}. In this line of research, deep neural networks (DNNs) are proposed to treat the MIMO channel matrix as an image, leveraging computer vision-inspired techniques for compression \cite{Guo20, Mashhadi21_distributed}. In \cite{Kim25}, diffusion model-based CSI compression is proposed by incorporating quantization into the generative model framework.
In addition, entropy coding is applied to reduce the average feedback rate by leveraging the non-uniform distribution of quantized features, at the cost of variable-length feedback \cite{Carpi23}.
Extensions to MU-MIMO have been explored through joint channel estimation and precoding \cite{Sohrabi21}, while alternative methods shift toward vector quantization rather than bit-based compression \cite{Jang22}. The authors of \cite{Sohrabi21, Jang22} further propose trainable downlink pilots adapted to channel distributions, focusing on pilot optimization rather than pilot reduction. Another line of research directly addresses pilot overhead reduction. In \cite{Park25}, channel-adaptive pilots are developed for TDD systems, enabling end-to-end optimization of channel acquisition and precoding with only a few pilots. Additionally, pilot pruning techniques based on fully connected (FC) layer-based designs are introduced to lower pilot transmission costs \cite{Mashhadi21_prun}. Despite these advances, a significant portion of CSI feedback, which comes from UEs that remain unscheduled, is wasted, suggesting that a selective feedback mechanism could further reduce unnecessary transmissions.

\subsection{Contributions and Organization}

This paper proposes a deep learning-based feedback reduction framework, \emph{self-nomination}, for MU-MIMO systems under average sum feedback rate constraints. Our main contributions are detailed below.

\textbf{Self-nomination approach for UE-side feedback reduction.}
A DNN for UE-driven feedback decisions is proposed, enabling each user to determine whether or not to send CSI to the BS. By selectively omitting feedback from users with a low likelihood of being scheduled, the self-nomination approach substantially lowers CSI overhead while maintaining system performance, and may even improve it by reducing interference in cases where scheduling fewer users is beneficial. Compared to prior methods that rely solely on SNR/SINR thresholding, the proposed approach can capture full spatial channel information. Simulation results show that a self-nomination strategy based on the complete channel vector outperforms decisions relying only on channel quality indicators (CQI), especially when the azimuth resolution is limited, as in uniform planar arrays (UPA) with few horizontal antennas, which leads to highly correlated user channels. In such cases, CQI-based methods fail to capture spatial congestion among UEs, resulting in suboptimal feedback decisions.
% especially when the spatial degrees of freedom (DoFs) are limited, such as in uniform planar arrays (UPA). 
Finally, self-nomination DNNs can be applied to diverse channel distributions, including realistic 3GPP models, and are not limited to Rayleigh fading.
%demonstrating greater adaptability than threshold-based methods that assume idealized Rayleigh fading. 
% By learning when feedback is truly beneficial, self-nomination reduces both power consumption at the UE and scheduling complexity at the BS.

\textbf{Training methods to handle non-differentiable decision-making and user scheduling.}
We formulate the feedback reduction problem as a weighted sum-rate maximization under an uplink sum feedback rate constraint and propose two complementary optimization strategies for training the self-nomination network. The first, a \textit{direct optimization-based approach}, adapted from our prior conference work, employs a primal-dual learning framework with a Lagrangian reformulation for unsupervised learning and handles non-differentiability via gradient approximations for both hard feedback decisions and user scheduling.
Building on this \mbox{unconstrained} formulation, we further develop a \textit{policy gradient-based approach} that models the feedback decision as a stochastic Bernoulli policy and optimizes it using policy gradients. By avoiding the need to differentiate through the scheduling step, this approach provides a more modular and general training procedure, compatible with various scheduling policies.
While the stochastic policy is flexible and general, the direct optimization approach naturally yields a deterministic feedback policy, which is more practical in systems requiring hard decisions. Together, the two methods provide both a practical baseline and a general solution.

\textbf{Extension to proportional-fair (PF) scheduling.}
The proposed self-nomination framework is extended to fairness-aware scheduling by incorporating a weighted sum-rate objective, where user weights are dynamically updated based on a PF algorithm. An additional input provides per-user weight information, enabling fairness-aware feedback decisions. 
To address the challenge of unknown weight distributions in PF scheduling, we adopt a simple yet effective sampling strategy during training, where PF weights are uniformly sampled from $[0,1]$, allowing the network to adapt to varying fairness scenarios and prioritize high-impact CSI feedback.
% A \blue{simple but effective} sampling strategy for PF weights is introduced for the training purpose, allowing the network to adapt to varying weight distributions and prioritize high-impact CSI feedback.
Simulation results show that self-nomination maintains fairness performance comparable to full feedback while reducing CSI overhead by approximately 65\%, based on a total of 100 users. This demonstrates its efficiency in balancing throughput and fairness with minimal feedback cost.

% \subsection{Paper Organization}
% The rest of this paper is organized as follows. Section~\ref{sec:system_model} presents the system model and problem formulation. Section~\ref{section:direct_optimization-based_approach} introduces the self-nomination framework, including direct optimization and policy gradient-based approaches. Section~\ref{sec:Self-Nomination for Proportional-Fair Scheduling} extends self-nomination to proportional-fair (PF) scheduling. Section~\ref{sec:performance_analysis} details the simulation setup and evaluation. Finally, Section~\ref{sec:conclusion} concludes the paper. 

% \subsection{Organization and Notation}
The rest of this paper is organized as follows. Section~\ref{sec:system_model} presents the system model and problem formulation. Section~\ref{section:direct_optimization-based_approach} introduces the self-nomination framework and direct optimization-based approach, while Section~\ref{sec:policy gradient-based approach} describes the policy gradient-based method. Section~\ref{sec:Self-Nomination for Proportional-Fair Scheduling} extends self-nomination to proportional-fair scheduling. Section~\ref{sec:performance_analysis} evaluates the proposed methods through simulations, and Section~\ref{sec:conclusion} concludes the paper.

\section{System Model}\label{sec:system_model}

We consider a MU-MIMO system, where the BS is equipped with $N$ antennas and each UE is equipped with a single antenna. Let $\bar{\mathcal{K}}$ denote the set of all UEs in the system. Each UE $k \in \bar{\mathcal{K}}$ observes its own downlink CSI, denoted by $\mathbf{h}_k \in \mathbb{C}^{N}$, and independently decides whether or not to feed back this CSI to the BS. We refer to this decision process as \emph{self-nomination}.

\subsection{Feedback Model at the UE Side} \label{subsec:Feedback Model at the UE Side}
To model the feedback decision at UE $k$, we use a DNN $f_{k}(\cdot;{\mathbf{\Theta}}_k) : \mathbf{h}_{k} \mapsto \{0,1\}$ parameterized by ${\mathbf{\Theta}}_k$. In particular, $f_{k}(\mathbf{h}_{k}) = 1$ indicates that UE $k$ feeds back its CSI, while $f_{k}(\mathbf{h}_{k}) = 0$ indicates it does not. We define the set of self-nominated UEs by $\mathcal{K}\triangleq\{k\in\bar{\mathcal{K}}:f_{k}(\mathbf{h}_{k}) = 1\}$. For simplicity, we assume that all UEs share the same DNN architecture and weight parameters, leading to $f(\cdot;{\mathbf{\Theta}}) = f_{1}(\cdot;{\mathbf{\Theta}}_1) = f_{2}(\cdot;{\mathbf{\Theta}}_2) = \cdots = f_{|\bar{\mathcal{K}}|}(\cdot;{\mathbf{\Theta}}_K).$

In practice, UE that decides to feed back its CSI would transmit $q(\mathbf{h}_k)$, a quantized version of the channel vector $\mathbf{h}_k$. However, to isolate the impact of self-nomination from the quantization procedure, we assume in this paper that the full, unquantized channel vector $\mathbf{h}_k$ is fed back.

\subsection{Scheduling and Precoding Model at the BS}
\label{subsec:scheduling_and_predocing_model}
After collecting CSI from all self-nominated UEs in the set $\mathcal{K}\subseteq \bar{\mathcal{K}}$, 
the BS schedules up to $M$ of them for downlink transmission. 
Formally, let $\mathcal{M} = g(\{{\mathbf{h}}_k\}_{k\in{\mathcal{K}}})$, or simply $g(\mathcal{K})$, where $g$ is a \emph{scheduling function} that outputs a subset $\mathcal{M} \subseteq \mathcal{K}$ with $\lvert \mathcal{M}\rvert \le M \le \lvert \mathcal{K}\rvert$.
In this work, we consider three specific scheduling methods.

\noindent
\textbf{Random Scheduling.}
\begin{equation*}
   g_{\mathrm{rand}}(\mathcal{K})\!=\!
   \begin{cases}
      \text{randomly choose $M$ UEs from $\mathcal{K}$},
      &\text{if }\lvert \mathcal{K}\rvert > M,\\[3pt]
      \mathcal{K}, &\text{otherwise}.
   \end{cases}
\end{equation*}

\noindent
\textbf{Opportunistic Scheduling.}
\begin{equation*}
   g_{\mathrm{opp}}(\mathcal{K})
   \;=\;
   \begin{cases}
      \displaystyle \mathop{\mathrm{arg~max}}_{\mathcal{S}\subseteq \mathcal{K},~
         \lvert \mathcal{S}\rvert = M} \sum_{k\in \mathcal{S}}\|\mathbf{h}_k\|_2^2,
      &\text{if } \lvert \mathcal{K}\rvert > M,\\[6pt]
      \mathcal{K}, &\text{otherwise}.
   \end{cases}
\end{equation*}

\noindent
\textbf{Proportional-Fair (PF) Scheduling.}
\begin{equation*}
   g_{\mathrm{PF}}(\mathcal{K}) \;=\;
  \mathop{\mathrm{arg~max}}_{\mathcal{M}\subseteq\mathcal{K},~
   \lvert \mathcal{M}\rvert \le M}
      U_{\mathrm{PF}}(\mathcal{M}),
\end{equation*}
where $U_{\mathrm{PF}}(\mathcal{M})$ denotes the PF utility function, which is detailed in Section~\ref{sec:Self-Nomination for Proportional-Fair Scheduling}. 
% \red{Jeff's Comment: Write down pros/cons of the 3 and why. Plus, mention why I don't consider the optimal scheduling.}

The three scheduling methods differ in complexity and objectives. Random scheduling is the simplest, ignoring channel quality, and generally yields the lowest sum-rate performance. Opportunistic scheduling enhances the sum-rate by selecting UEs with the largest channel gains, i.e., CQI, but may compromise fairness. PF scheduling balances throughput and fairness by incorporating long-term rate information through dynamically updated weights. In this work, these scheduling methods are integrated into our proposed framework. We avoid more complex scheduling, such as exhaustive search, as it is impractical for DNN training, where scheduling must be performed for every mini-batch sample across numerous epochs, leading to excessive computational overhead.

\begin{figure}[t]
\begin{center}
\includegraphics[width=\linewidth]
{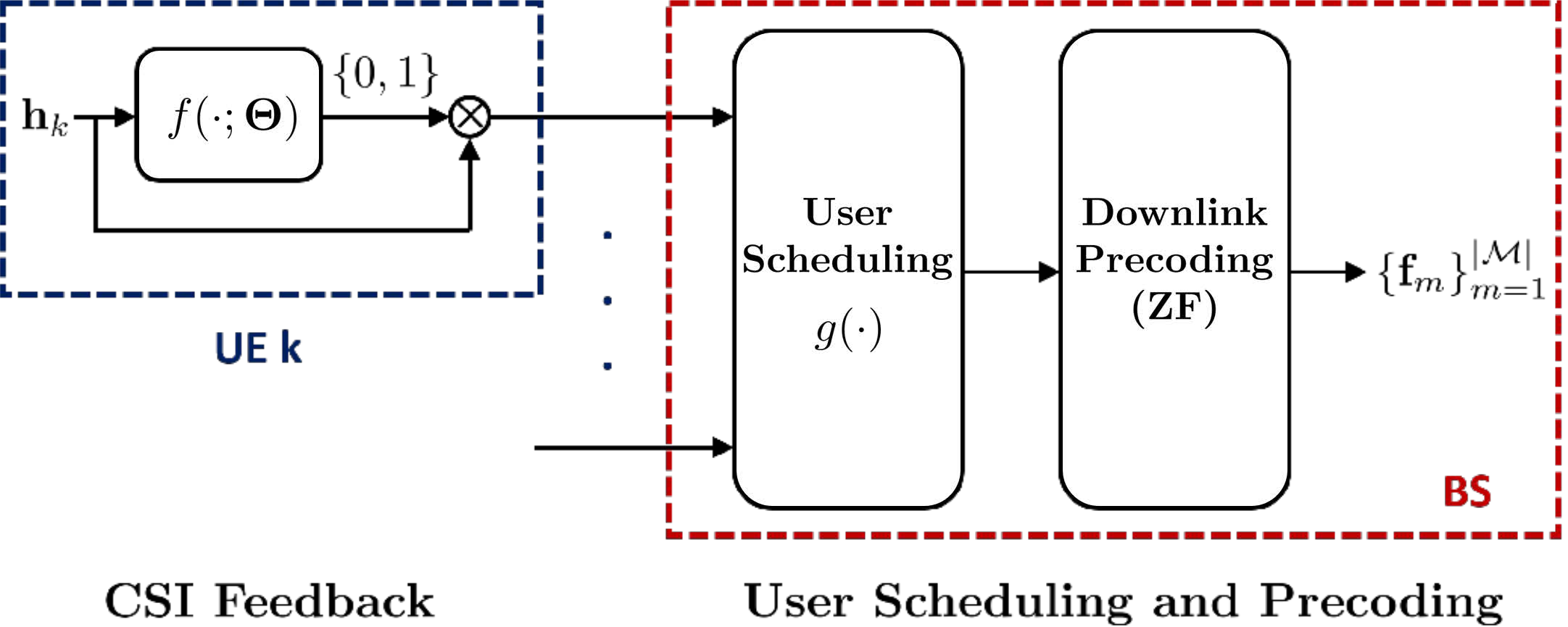}
\end{center}
\vspace{-3mm}
\caption{The decentralized self-nominating system.}
\label{figure:system_model}
\end{figure}

Once the BS forms the scheduled set $\mathcal{M} = g(\mathcal{K})$, 
it applies zero-forcing (ZF) precoding to serve the UEs in $\mathcal{M}$. We adopt ZF precoding as it is simple and differentiable, and asymptotically achieves optimal performance in the high SNR regime.
Let $\mathbf{H} \triangleq \bigl[\mathbf{h}_1,\ldots,\mathbf{h}_{\lvert \mathcal{M}\rvert}\bigr]^H \in\mathbb{C}^{\lvert \mathcal{M}\rvert\times N}$ be the aggregated channel matrix of those scheduled UEs. 
The \emph{unnormalized} ZF precoder is
\begin{equation}
   \widetilde{\mathbf{F}}=\mathbf{H}^H \bigl(\mathbf{H}\mathbf{H}^H\bigr)^{-1},
\end{equation}
whose $k$-th column is denoted by $\widetilde{\mathbf{f}}_{k}$. 
To satisfy the total power $P$ and ensure equal power allocation among the $\lvert \mathcal{M}\rvert$ scheduled UEs, each column is normalized as 
$\mathbf{f}_{k} = \sqrt{\frac{P}{|\mathcal{M}|}}
{\widetilde{\mathbf{f}}_{k}}
/{\|\widetilde{\mathbf{f}}_{k}\|_2}.$
Hence, the final precoding matrix is
$\mathbf{F} = [\mathbf{f}_1,\dots,\mathbf{f}_{\lvert \mathcal{M}\rvert}]$, 
and the transmit signal is 
$\mathbf{x} = \mathbf{F}\,\mathbf{s}$,
where $\mathbf{s}\in\mathbb{C}^{\lvert \mathcal{M}\rvert}$ satisfies 
$\mathbb{E}[\mathbf{s}\mathbf{s}^H] = \mathbf{I}_{\lvert \mathcal{M}\rvert}$. Here, $\mathbf{I}_{\lvert \mathcal{M}\rvert}$ denotes the $\lvert \mathcal{M}\rvert \times \lvert \mathcal{M}\rvert$ identity matrix.
Fig.~\ref{figure:system_model} summarizes the system model.

\vspace{-2mm}

\subsection{Problem Formulation}\label{subsec:Problem Formulation}
A common issue with existing feedback reduction algorithms is that the sum feedback rate increases with the number of UEs. However, in practical systems, the total uplink feedback capacity is strictly limited due to resource-constrained control channels, such as the physical uplink control channel (PUCCH) in LTE and 5G NR, allowing only a small number of slots for CSI feedback per transmission.

To account for this uplink feedback capacity constraint, we adopt the average sum feedback rate formulation in \cite{Huang07}, given by
\begin{align}
B \cdot \mathbb{E}\biggl[\sum_{k \in \bar{\mathcal{K}}} f\bigl(\mathbf{h}_k\bigr)\biggr] \;\le\; C_{\mathrm{FB}},
\end{align}
where $B$ is the number of bits allocated per UE, and $C_{\mathrm{FB}}$ denotes the total feedback rate budget. This constraint ensures that the expected total number of feedback bits per transmission does not exceed $C_{\mathrm{FB}}$. 
While $B$ could be optimized jointly with $f(\cdot)$, we fix it to a constant to focus on the effect of self-nomination on UE scheduling and precoding.
% Note that while $B$ could, in principle, be optimized jointly with the feedback decision function $f(\cdot)$, we fix it to a constant to focus on the effect of self-nomination on UE scheduling and precoding.
For convenience, we define the dimensionless parameter $N_{\mathrm{FB}} \triangleq C_{\mathrm{FB}} / B$, which represents the average limit on the number of UEs that can feed back under this constraint.

\begin{figure*}[t]
\begin{center}
\includegraphics[width=5in]
{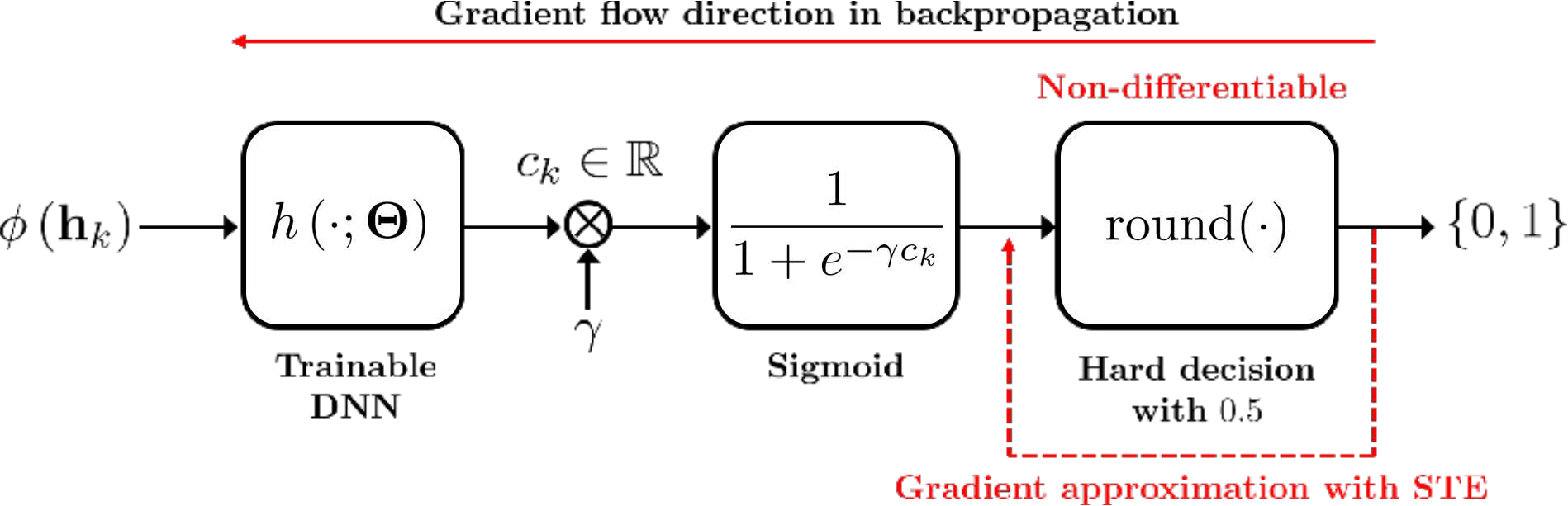}
\end{center}
\caption{Overall architecture of the proposed self-nomination network $f(\cdot;\mathbf{\Theta})$ for the direct optimization-based approach.}
\label{fig:proposed_DNN_for_DO}
\end{figure*}

Under the feedback constraint, the optimization problem can be expressed as
\begin{subequations}\label{eqn:problem_formulation}
\begin{align}
% \mathop {{\rm{max}}}\limits_{f\left(  \cdot ;\mathbf{\Theta} \right) }~~~~ &\mathbb{E}\left[ {\sum\limits_{m \in {\cal M}} {{w_m}{{\log }_2} \left( {1 + \mathrm{SINR}_m} \right)} } \right]\\
\mathop {{\rm{max}}}\limits_{f\left(  \cdot ;\mathbf{\Theta} \right) }~~~~ &\mathbb{E}\left[ {\sum\limits_{m \in {\cal M}} {{w_m}R_m} } \right]\\
\text{subject to}~~
&\mathbb{E}\left[ \sum\limits_{k \in \bar {\mathcal{K}}} {f\left( \mathbf{h}_k;\mathbf{\Theta} \right)} \right] \le N_{\mathrm{FB}},
\end{align}
% \tag{P1} % Labeling as (P1)
\end{subequations} %\unskip
where $w_m$ is the weight of the $m$-th scheduled user, and the spectral efficiency $R_m$ is defined by ${{\log }_2} \left( {1 + \mathrm{SINR}_m} \right)$. Note that $\boldsymbol{\Theta}$ represents the DNN parameters of the UE-side binary decision function $f(\cdot;\boldsymbol{\Theta})$, which we aim to optimize by maximizing the objective in \eqref{eqn:problem_formulation}.
% , which we optimize in an unsupervised manner to maximize the above objective. 
As for the weights, we first focus on the special case of uniform weights $(w_m = 1)$, which reduces to a sum-rate maximization problem, and later generalize to PF weights in Section~\ref{sec:Self-Nomination for Proportional-Fair Scheduling}.

For each scheduled UE $m$ with its corresponding precoder $\mathbf{f}_m$, 
the SINR is given by
\begin{align}
\mathrm{SINR}_m = \frac{\left| \mathbf{h}_m^H \mathbf{f}_m \right|^2}{\sigma^2 + \sum\limits_{n \in \mathcal{M}, n \neq m} \left| \mathbf{h}_m^H \mathbf{f}_n \right|^2},
\label{eqn:sinr_original}
\end{align}
where $\sigma^2$ represents the noise power. Since ZF precoding ideally nullifies multiuser interference for well-conditioned channels, the interference term in \eqref{eqn:sinr_original} may be negligible, and the SINR reduces to
\begin{align}
\mathrm{SINR}_m \approx \frac{\bigl|\mathbf{h}_m^H \mathbf{f}_m\bigr|^2}{\sigma^2}=\mathrm{SNR}_m.
\label{eqn:sinr_reduced}
\end{align}
However, in practice, when the channel matrix $\mathbf{H}$ is ill-conditioned, residual interference may persist. Even without significant interference, $\mathrm{SINR}_m$ is small if the effective channel gain $\bigl|\mathbf{h}_m^H \mathbf{f}_m\bigr|^2$ is small.

Suboptimal selection of $\mathcal{M}$ can degrade $\mathrm{SINR}_m$.
First, if the scheduled UEs have highly correlated channels, the resulting ZF precoders become poorly aligned with each individual $\mathbf{h}_m$, reducing $\bigl|\mathbf{h}_m^H \mathbf{f}_m\bigr|^2$. 
Second, scheduling users with inherently weak channels (i.e., low channel gain) will limit their achievable $\mathrm{SINR}$, even if interference is largely mitigated. Thus, carefully choosing $\mathcal{M}$ is critical to ensure stronger and less-correlated channels. Consequently, random scheduling and opportunistic scheduling are obviously suboptimal.

To address these limitations, the self-nomination step restricts feedback to the most promising UEs, even though each UE only has access to its local CSI. In this approach, each UE’s DNN decides whether to feed back based on the statistics of its own channel and the anticipated impact of channel correlation with other users. By coordinating self-nomination with a fixed scheduling procedure, the system can prioritize users that offer higher effective SINR, thereby enhancing overall performance even under a suboptimal or complexity-constrained BS-side scheduler.

% \begin{figure}[t]
% \centering
% \subfigure[Overall architecture of the proposed self-nomination network $f(\cdot;\mathbf{\Theta})$.]{
%     \includegraphics[width=\linewidth]
%     % \includegraphics[width=5in]
%     {figure/figure_proposed_new_ver2.png}
%     \label{fig:proposed_overall}}
%     \vspace{10pt}
% \subfigure[The trainable DNN $g(\cdot;\mathbf{\Theta})$ architecture when $\protect\phi(\mathbf{h}_k) = \protect\mathbf{h}_k$.]{
%     \includegraphics[width=0.8\linewidth]
%     % \includegraphics[width=4in]
%     {figure/figure_proposed_g_fullchannel.png}
%     \label{fig:proposed_g_full_channel}}
%     % \vspace{5pt}    
% \subfigure[The trainable DNN $g(\cdot;\mathbf{\Theta})$ architecture when $\protect\phi(\mathbf{h}_k) = \protect\|\mathbf{h}_k\|_2$.]{
%     \includegraphics[width=0.7\linewidth]
%     % \includegraphics[width=3.4in]
%     {figure/figure_proposed_g_cqi.png}
%     \label{fig:proposed_g_cqi}}
% \caption{Illustration of the proposed self-nomination network and its architectures.}
% \label{figure:proposed_DNN}
% \end{figure}

\section{Direct Optimization-Based Approach} \label{section:direct_optimization-based_approach}
In this section, we detail the proposed binary feedback decision DNN, referred to as the \emph{self-nominating DNN}, for the case of random scheduling and ZF precoding. We then explain how to perform direct optimization via gradient descent by formulating the Lagrangian of \eqref{eqn:problem_formulation}.

% \begin{figure*}[tb]
% \begin{align}
% \mathbf{\Theta}_{n+1} \leftarrow \mathbf{\Theta}_{n} - \frac{\alpha_\mathrm{p}}{|A|} \sum_{a \in A} \left( \nabla_{\mathbf{\Theta}} \left[ \sum_{m \in \mathcal{M}_a} w_m R_m - \lambda_{n} \left( \sum_{k \in \bar{\mathcal{K}}_a} f\left( \mathbf{h}_k ; \mathbf{\Theta}_{n}\right) - N_{\mathrm{FB}} \right) \right] \right)
% \label{eqn:primal_update}
% \end{align}
% \end{figure*}

\subsection{DNN Architecture and Gradient Approximation}
\label{subsec:DNN Architecture and Gradient Approximation}

Fig.~\ref{fig:proposed_DNN_for_DO} shows the overall architecture of our self-nomination network, $f(\cdot; \mathbf{\Theta})$, which outputs a binary decision indicating whether UE $k$ should nominate itself for feedback.
The network has two main parts: a trainable DNN ${h}(\cdot; \mathbf{\Theta})$, parameterized by $\mathbf{\Theta}$, and a subsequent binarization module.

The input $\phi(\mathbf{h}_k)$ for UE $k$ can be either the full channel vector $\mathbf{h}_k$ or a CQI, specifically the $\ell_{2}$-norm $\|\mathbf{h}_k\|_{2}$. In the former case ($\phi(\mathbf{h}_k) = \mathbf{h}_k$), one-dimensional (1D) convolutional layers first extract spatial-frequency features, followed by FC layers with batch normalization and a final $\tanh$ activation, where $\tanh$ denotes the hyperbolic tangent function.
% , as illustrated in Fig.~\ref{fig:proposed_g_full_channel}. 
If $\phi(\mathbf{h}_k) = \|\mathbf{h}_k\|_{2}$, the convolutional stages are omitted, and the norm value is fed directly into the same FC layers, including batch normalization and the final $\tanh$ activation.
% the FC layers.
% shown in Fig.~\ref{fig:proposed_g_cqi}.

In both cases, the DNN produces a scalar output $c_k = h(\mathbf{h}_k;\mathbf{\Theta}) \mathrm{~or~} h(\|\mathbf{h}_k\|_2;\mathbf{\Theta}) \in \mathbb{R}$.
A sharpness parameter $\gamma$ scales $c_k$ before it is passed to a logistic function 
$\mathrm{sigm}(x) = 1/(1 + e^{-x})$, giving $\mathrm{sigm}(\gamma \, c_k) \in [0, 1].$
A hard threshold at $0.5$ then yields $\mathrm{round}(\mathrm{sigm}(\gamma \, c_k))$.
% \begin{equation}\label{eqn:self_nomination_hard_decision}
% f(\mathbf{h}_k; \mathbf{\Theta}) 
% \,=\,
% \begin{cases}
% 1, & \text{if } \mathrm{sigm}(\gamma \, c_k) \,\ge\, 0.5,\\
% 0, & \text{otherwise}.
% \end{cases}
% \end{equation}
This final binary decision indicates whether the UE nominates itself for feedback. 
Higher values of $\gamma$ make the sigmoid transition sharper, 
leading to a more distinct decision boundary around $c_k = 0$.

However, the proposed DNN architecture faces two major sources of non-differentiability that block gradient propagation. First, the binary decision $f(\mathbf{h}_k;\mathbf{\Theta})$ is obtained by thresholding the continuous sigmoid output $\mathrm{sigm}(\gamma\,c_k)$, whose unit step function has zero or undefined gradient. We address this using a \emph{straight-through estimator} (STE) \cite{Bengio13}. In the forward pass, the hard threshold is applied as originally defined, while in the backward pass, $\partial f_k/\partial c_k$ is approximated by ignoring the hard-thresholding and directly using the derivative of the sigmoid function, given by
\begin{equation}
   \frac{\partial f_k}{\partial c_k}
   \approx
   \frac{\partial }{\partial c_k}\mathrm{sigm}(\gamma\,c_k)
   =\gamma\,\mathrm{sigm}(\gamma\,c_k)\bigl(1-\mathrm{sigm}(\gamma\,c_k)\bigr).
\end{equation}
This approximation allows gradients to propagate through the binary decision.

% Second, the scheduling functions $g(\cdot)$ are also non-differentiable with respect to $\mathbf{h}_k, \forall k \in \mathcal{K}$. Here, we apply \emph{selective gradient propagation}, whereby only scheduled UEs contribute to the gradient.
% For the scheduled UE set $\mathcal{M}\subseteq\mathcal{K}$, the loss or reward gradient is accumulated only over UEs $m \in \mathcal{M}$. This bypasses the need to compute gradients of a discrete sampling procedure and still captures the effect of each UE’s decision on the final performance.

Second, the scheduling function $g(\cdot)$ is non-differentiable with respect to each $\mathbf{h}_k$, $k\in\mathcal{K}$. 
We address this by \emph{selective gradient propagation}, 
which treats $g(\cdot)$ as an identity mapping for scheduled users 
and zero for non-scheduled users in the backward pass. 
% Specifically, if $\mathcal{M} = g(\{{\mathbf{h}}_k\}_{k\in{\mathcal{K}}})$ is the scheduled set, 
% we approximate the gradient as 
% \begin{equation}
%    \frac{\partial g(\{{\mathbf{h}}_k\}_{k\in{\mathcal{K}}})}{\partial \mathbf{h}_m}
%    \;=\;
%    \begin{cases}
%       \mathbf{I}, & m \in \mathcal{M}, \\
%       \mathbf{0}, & m \notin \mathcal{M}.
%    \end{cases}
% \end{equation}
Thus, when backpropagating a loss that depends on $g(\cdot)$ only through $\mathcal{M}$, 
the partial derivatives with respect to $\{\mathbf{h}_k : k \notin \mathcal{M}\}$ vanish. 
This ensures that only the scheduled UEs contribute to the gradient, 
bypassing the non-differentiable selection step while still capturing the impact of self-nomination on system performance.

\begin{figure*}[t]
\begin{center}
\includegraphics[width=5in]
{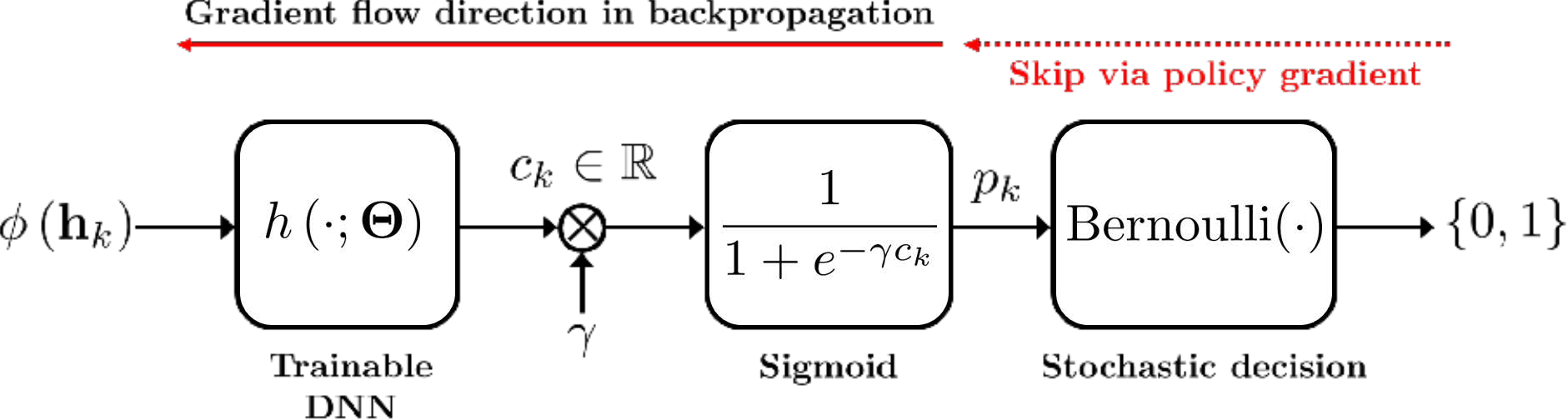}
\end{center}
\caption{Overall architecture of the proposed self-nomination network $f(\cdot;\mathbf{\Theta})$ for the policy gradient-based approach.}
\label{fig:proposed_DNN_for_PG}
\end{figure*}

\subsection{Lagrangian Optimization and Training Strategy}
\label{subsec:Lagrangian Optimization and Training Strategy}
The absence of labeled optimal feedback decisions necessitates an unsupervised learning approach for maximizing the weighted sum-rate. However, the constraint in \eqref{eqn:problem_formulation} prevents the direct application of unsupervised learning.
% Our optimization problem, as formulated in \eqref{eqn:problem_formulation}, includes a constraint that prevents the direct application of unsupervised learning.
To address this challenge, we transform the constrained optimization problem into an unconstrained one by incorporating the constraint into the objective function using a Lagrangian formulation. This approach introduces a dual variable corresponding to the constraint, motivating us to employ a primal-dual learning method \cite{Eisen19, Lee19}, which alternates between updating the primal variables $\mathbf{\Theta}$ and the dual variable $\lambda$.

The Lagrangian function for our problem can be defined as  
\begin{align}
\mathcal{L}(\mathbf{\Theta}, \lambda) = &\mathbb{E}\left[ {\sum \limits_{m \in {\cal M}} {{w_m}R_m} } \right] \nonumber\\
&+ \lambda \left(\mathbb{E}\left[ \sum\limits_{k \in \bar {\mathcal{K}}} {f\left( \mathbf{h}_k ; \mathbf{\Theta}\right)} \right] - N_{\mathrm{FB}} \right),
\label{eqn:lagrangian}
\end{align}
where $\mathbf{\Theta}$ denotes the parameters of the UE self-nomination network, and $\lambda$ is the dual variable associated with the constraint. Note that $\mathrm{SINR}_m$ is influenced by the UE self-nomination network $f( \cdot ; \mathbf{\Theta} )$, as the pool of self-nominated UEs affects the subsequent UE scheduling and ZF precoding.

In the primal-dual learning framework, we alternately update the primal variables $\mathbf{\Theta}$ and the dual variable $\lambda$. The primal variables are updated via stochastic gradient descent (SGD), keeping the dual variable $\lambda$ fixed. The update rule for the primal variables is given by
\begin{align} 
\mathbf{\Theta}_{n+1} \leftarrow \mathbf{\Theta}_{n} - \alpha_{\mathrm{p}} \nabla_{\mathbf{\Theta}} \mathcal{L}(\mathbf{\Theta}_{n}, \lambda_{n}), 
\label{eqn:primal_update_simplified} 
\end{align} 
where $\alpha_{\mathrm{p}}$ is the learning rate for the primal variables, and $\nabla_{\mathbf{\Theta}} \mathcal{L}$ denotes the gradient of the Lagrangian with respect to ${\mathbf{\Theta}}$. When computing $\nabla_{\mathbf{\Theta}} \mathcal{L}$, we apply STE and selective gradient propagation as described in Section \ref{subsec:DNN Architecture and Gradient Approximation}. The expectation in the Lagrangian is then approximated using a mini-batch drawn from a subset of training samples.

After updating the primal variables, we update the dual variable $\lambda$ using the subgradient method, keeping the primal variables fixed. The update rule for the dual variable is
\begin{align} 
\lambda_{n+1} \leftarrow \left[ \lambda_{n} + \alpha_{\mathrm{d}} \nabla_{\lambda} \mathcal{L}(\mathbf{\Theta}_{n+1}, \lambda_{n}) \right]_+, \label{eqn:dual_update_general} 
\end{align} 
where $\alpha_{\mathrm{d}}$ is the learning rate for the dual variable, and $[\cdot]_+$ denotes the projection onto the non-negative orthant to ensure that $\lambda \geq 0$. Here, we choose the gradient $\nabla_{\lambda} \mathcal{L}$ as our subgradient, ensuring that the update direction for $\lambda$ directly addresses the constraint violation. The gradient of the Lagrangian with respect to $\lambda$ is given by
\begin{align} 
\nabla_{\lambda} \mathcal{L}({\mathbf{\Theta}}, \lambda) = \mathbb{E}\left[ \sum_{k \in \bar{\mathcal{K}}} f( \mathbf{h}_k ; \mathbf{\Theta} ) \right] - N_{\mathrm{FB}}.
\end{align} 
Similar to the primal update, we approximate the expectation using a mini-batch during training.

In summary, our training strategy involves alternately updating the primal variables $\mathbf{\Theta}$ and the dual variable $\lambda$ using stochastic optimization methods.
Specifically, the updates are performed on a mini-batch basis, where the contribution of each sample is averaged within the batch. 
This primal-dual approach allows us to effectively handle the constraints in our optimization problem, ensuring convergence to a solution that satisfies the original constraints while optimizing the objective function.

\section{Policy Gradient-Based Approach}\label{sec:policy gradient-based approach}

In Section \ref{section:direct_optimization-based_approach}, we introduced gradient approximation techniques to optimize the self-nomination DNN. However, these methods rely on heuristic approximations that introduce errors during backpropagation, potentially leading to suboptimal solutions. To address this limitation, we propose a policy gradient-based approach, which reformulates the self-nomination decision as a stochastic process, avoiding direct gradient computation through non-differentiable operations.

\subsection{Stochastic Decision Policy}

To eliminate the reliance on heuristic gradient approximations using STE, we replace the deterministic thresholding operation with a stochastic decision policy. This policy allows gradient-based optimization through probabilistic sampling while preserving the ability to learn deterministic behavior.

\begin{prop} \label{prop:1}
A stochastic policy generalizes a deterministic policy, as any deterministic policy can be expressed as a stochastic policy with a degenerate distribution concentrated on a single action.
\end{prop} 

% \begin{proof}
% A deterministic policy selects a fixed action $a^*$ for a given state $s$, which can be viewed as a special case of a stochastic policy with a degenerate distribution: 
% $p(a \mid s) = 1$ if $a = a^*$ and $0$ otherwise. Thus, every deterministic policy can be expressed as a stochastic policy with all probability mass concentrated on a single action.
% \end{proof}

% The proof is omitted as it follows directly from the definition of stochastic policies and degenerate distributions.

Motivated by Proposition~\ref{prop:1}, we model the self-nomination decision as a Bernoulli random variable:
\begin{equation}\label{eqn:def_f_pg}
    f(\mathbf{h}_k; \mathbf{\Theta}) \sim \mathrm{Bernoulli}(p_k),
\end{equation}
where the probability $p_k$ is given by
\begin{equation}\label{eqn:def_p_k}
    p_k = \mathrm{sigm}(\gamma \, c_k) = \frac{1}{1 + e^{-\gamma \, c_k}}.
\end{equation}
Fig.~\ref{fig:proposed_DNN_for_PG} depicts the modified DNN architecture for the policy gradient-based approach, which incorporates Bernoulli($p_k$) sampling to implement the stochastic policy, in contrast to the deterministic hard-thresholding used in Fig.~\ref{fig:proposed_DNN_for_DO}.
As in Fig.~\ref{fig:proposed_DNN_for_DO}, $c_k = h(\mathbf{h}_k; \mathbf{\Theta})$ is the scalar output of the DNN, and $\gamma$ controls the sharpness of the probability transition. During training, the DNN tends to push $c_k$ toward large positive or negative values in most cases, driving $p_k$ to 0 or 1.
% and recovering deterministic behavior as a special case of the stochastic policy. 
This results in near-deterministic behavior, making the deterministic policy a special case of the stochastic policy.
% Importantly, the learned distribution of $c_k$ tends to grow in magnitude except in some regions, pushing $p_k = \mathrm{sigm}(\gamma \, c_k)$ toward 0 or 1. This effectively makes $f(\mathbf{h}_k; \Theta)$ a deterministic function of the input, even for a fixed $\gamma$, aligning with Proposition~\ref{prop:1}.

Denoting by $\pi(\mathbf{a} \mid \{{\mathbf{h}_k\}_{k \in \bar{\mathcal{K}}}}; \mathbf{\Theta})$ the global stochastic policy governing feedback decisions for all UEs, we exploit the decentralized structure of the system to factorize this global policy into local policies, given by
\begin{equation}
\pi(\mathbf{a} \mid \{{\mathbf{h}_k\}_{k \in \bar{\mathcal{K}}}}; \mathbf{\Theta}) = \prod_{k \in \bar{\mathcal{K}}} \pi(a_k \mid \mathbf{h}_k; \mathbf{\Theta}),
\end{equation}
where $\mathbf{a} = \{a_k\}_{k \in \bar{\mathcal{K}}}$ is the set of binary feedback decisions ($a_k \in \{0,1\}$), and each local policy $\pi(a_k \mid \mathbf{h}_k; \mathbf{\Theta})$ depends only on the locally observed CSI $\mathbf{h}_k$. 
Specifically, each UE’s local policy follows a Bernoulli distribution parameterized by the self-nomination probability $p_k$, defined in \eqref{eqn:def_p_k}, such that $\pi(a_k \mathrel{=} 1 \mid \mathbf{h}_k; \mathbf{\Theta})= p_k$.
Thus, the global policy becomes
\begin{equation}\label{eqn:product_policies}
\pi(\mathbf{a} \mid \{{\mathbf{h}_k}\}_{k \in \bar{\mathcal{K}}}; \mathbf{\Theta}) = \prod_{k \in \bar{\mathcal{K}}} p_k^{a_k} (1 - p_k)^{1 - a_k}.
\end{equation}
The global policy is thus obtained by multiplying the probabilities of independent local decisions, each determined by local CSI and its corresponding self-nomination probability.
% While more elaborate joint policies could, in principle, be designed based on global CSI, such approaches are impractical due to excessive signaling overhead. In contrast, the proposed decentralized policy structure enables scalable operation, as the complexity of the self-nomination decision does not grow with the number of UEs. The use of a shared DNN architecture further ensures consistent and lightweight decision-making across all UEs.

\subsection{Policy Gradient for Self-Nomination}

We now derive the policy gradient method for optimizing the stochastic self-nomination policy. Building on the Lagrangian formulation in \eqref{eqn:lagrangian}, we treat the self-nomination decision as a stochastic policy and compute gradients using the log-derivative trick \cite{Sutton00}. The objective is defined as
\begin{equation}
    J(\mathbf{\Theta}) = \mathbb{E}_{\mathbf{a} \sim \pi(\cdot | \{\mathbf{h}_k\}; \mathbf{\Theta})} \left[ \mathcal{L}(\{\mathbf{h}_k\}, \mathbf{a}, \lambda) \right],
    \label{eqn:lagrangian_for_pg}
\end{equation}
where the expectation is taken over the stochastic policy $\pi(\mathbf{a} \mid \{{\mathbf{h}_k\}_{k \in \bar{\mathcal{K}}}}; \mathbf{\Theta})$, which governs the self-nomination decisions of all UEs. For notational convenience, we let $\{\mathbf{h}_k\}$ denote the set of channel states for all UEs $\{\mathbf{h}_k\}_{k \in \bar{\mathcal{K}}}$. 
We explicitly include the channel states $\{\mathbf{h}_k\}$ as arguments of the Lagrangian to maintain notational consistency with the policy, which is conditioned on the same set.
% $\{\mathbf{h}_k\}$.

Unlike the original formulation in \eqref{eqn:lagrangian}, where the Lagrangian $\mathcal{L}$ is directly parameterized by $\mathbf{\Theta}$, the expression in \eqref{eqn:lagrangian_for_pg} treats it as a function of the action $\mathbf{a}$ and the user channels, with no explicit dependence on $\mathbf{\Theta}$. Instead, the influence of $\mathbf{\Theta}$ on the objective arises through the policy $\pi(\mathbf{a} \mid \{\mathbf{h}_k\}; \mathbf{\Theta})$, which defines the distribution over actions. That is, $\mathbf{\Theta}$ affects the optimization objective only in expectation, through how it shapes the distribution of $\mathbf{a}$. For each UE $k$, the action $a_k$ is sampled from a $\mathrm{Bernoulli}(p_k)$ distribution, where $p_k$ is defined in \eqref{eqn:def_p_k}.
% , and the Lagrangian is then evaluated as a deterministic function of the sampled action and fixed channels.

Although $\mathbf{a}$ is sampled from the policy parameterized by $\mathbf{\Theta}$, the Lagrangian $\mathcal{L}(\{\mathbf{h}_k\}, \mathbf{a}, \lambda)$ is evaluated at a fixed realization of $\mathbf{a}$ and thus treated as independent of $\mathbf{\Theta}$. This allows the use of the log-derivative trick to compute gradients of the expected objective $J(\mathbf{\Theta})$ with respect to $\mathbf{\Theta}$. During this step, $\lambda$ is held fixed, so we write the Lagrangian as $\mathcal{L}(\{\mathbf{h}_k\}, \mathbf{a})$. The dual variable $\lambda$ is then updated separately in an alternating fashion, as described in Section~\ref{subsec:Lagrangian Optimization and Training Strategy}.

The following result characterizes the gradient of the expected Lagrangian with respect to the policy parameters during the primal update, where the dual variable $\lambda$ is held fixed and omitted from the notation.
\begin{prop}\label{prop:policy_gradient}
For a stochastic policy $\pi(\mathbf{a} \mid \{\mathbf{h}_k\}; \mathbf{\Theta})$ with an associated Lagrangian $\mathcal{L}(\{\mathbf{h}_k\},\mathbf{a})$, the gradient of the expected Lagrangian with respect to the policy parameters satisfies
\begin{align}
    \nabla_{\mathbf{\Theta}} \mathbb{E}_{\pi} &\left[ \mathcal{L}(\{\mathbf{h}_k\},\mathbf{a}) \right] \nonumber\\
    &= \mathbb{E}_{\pi} \left[ \left(\nabla_{\mathbf{\Theta}} \log \pi(\mathbf{a} \mid \{\mathbf{h}_k\}; \mathbf{\Theta})\right) \cdot \mathcal{L}(\{\mathbf{h}_k\}, \mathbf{a}) \right].
\end{align}
\end{prop}
\textbf{Proof:} The result follows directly from the log-derivative trick \cite{Sutton00} in stochastic optimization. A complete proof is provided in Appendix~\ref{app:proof_prop_policy_gradient}.

Unlike the standard policy gradient method in \cite{Sutton00}, which optimizes policies over multiple time steps for sequential decision-making, our approach considers a single-decision step for each self-nomination, as CSI feedback is determined independently under a block fading assumption.

To apply this result, we approximate the expectation using mini-batches of channel states and self-nomination decisions. For a mini-batch $\mathcal{A}$ of UEs, the gradient is approximated as
\begin{align}
    \nabla_{\mathbf{\Theta}} &\mathbb{E}_{\pi} \left[ \mathcal{L}(\{\mathbf{h}_k\},\mathbf{a}) \right] \nonumber\\
    \approx &\frac{1}{\lvert \mathcal{A} \rvert} \sum_{k \in \mathcal{A}} \left[ \left(\nabla_{\mathbf{\Theta}} \log \pi(\mathbf{a} \mid \{\mathbf{h}_k\}; \mathbf{\Theta})\right) \cdot \mathcal{L}(\{\mathbf{h}_k\},\mathbf{a}) \right].
\end{align}
% where $a_k \in \{0,1\}$ is the binary feedback decision for UE $k$.
For the Bernoulli self-nomination policy, the log probability of the local decision $a_k \in\{0,1\}$ is given by
\begin{equation}
    \log \pi(a_k \mid \mathbf{h}_k; \mathbf{\Theta}) = a_k \log p_k + (1 - a_k) \log (1 - p_k).
\end{equation}
% where $p_k = \mathrm{sigm}(\gamma \, c_k)$ is the probability of self-nomination, and $c_k = g(\mathbf{h}_k; \mathbf{\Theta})$ is the scalar output of the DNN. 
From \eqref{eqn:product_policies}, the global log probability is 
\begin{equation}
    \log \pi(\mathbf{a} \mid \{{\mathbf{h}_k}\}; \mathbf{\Theta}) = \sum_{k \in \bar{\mathcal{K}}} \left[a_k \log p_k + (1 \!- \!a_k) \log (1 - p_k)\right],
\end{equation}
and the gradient of the log probability with respect to $\mathbf{\Theta}$ is computed as
\begin{equation}\label{eqn:grad_global_log_prob}
    \nabla_{\mathbf{\Theta}} \log \pi(\mathbf{a} \mid \{{\mathbf{h}_k}\}; \mathbf{\Theta}) =  \sum_{k \in \bar{\mathcal{K}}} \left( \frac{a_k}{p_k} - \frac{1 - a_k}{1 - p_k} \right) \nabla_{\mathbf{\Theta}} p_k.
\end{equation}
Using \eqref{eqn:def_p_k}, applying the chain rule yields
\begin{equation}    
   \nabla_{\mathbf{\Theta}} p_k 
   \;=\; 
   \gamma \, p_k \,\bigl(1 - p_k\bigr)\,\nabla_{\mathbf{\Theta}} c_k,
\end{equation}
where $c_k = h(\mathbf{h}_k;\mathbf{\Theta})$ and $\nabla_{\mathbf{\Theta}} c_k$ is obtained via standard backpropagation through the DNN.  
% Substituting this expression for $\nabla_{\mathbf{\Theta}} p_k$ into \eqref{eqn:grad_global_log_prob} provides the global log-probability gradient in terms of $\mathbf{\Theta}$, enabling end-to-end training of the stochastic self-nomination policy.

Plugging \eqref{eqn:grad_global_log_prob} into Proposition~\ref{prop:policy_gradient}, we obtain an update rule for the primal variables $\mathbf{\Theta}$ based on stochastic gradient ascent:
\begin{equation}
    \mathbf{\Theta}_{n+1} \leftarrow \mathbf{\Theta}_{n} + \alpha_{\mathrm{p}} \nabla_{\mathbf{\Theta}_n} J(\mathbf{\Theta}_n),
\end{equation}
where $\alpha_{\mathrm{p}}$ is the primal learning rate. In parallel, the dual variable $\lambda$ is updated to enforce the feedback constraint using \eqref{eqn:dual_update_general}.

This approach eliminates the need for gradient approximations in the STE and directly optimizes the self-nomination policy while respecting the feedback constraint through the Lagrangian framework.

\section{Self-Nomination for Proportional-Fair Scheduling}\label{sec:Self-Nomination for Proportional-Fair Scheduling}

The proposed self-nomination framework can also be extended to scenarios where fairness is considered in user scheduling. To this end, we first reformulate the problem using a PF scheduling objective and then modify the self-nomination network accordingly to accommodate this fairness criterion.

\subsection{PF Scheduling via Weighted Sum-Rate Maximization}
To balance system throughput and user fairness, we adopt PF scheduling in the context of weighted sum-rate maximization, where user weights dynamically adjust based on past average spectral efficiencies \cite{Cui19}, \cite[Section~7.6.1]{Heath18}. 

Let $\bar{R}_k[t]$ denote the running average of UE $k$’s spectral efficiency at time $t$, defined as
\begin{equation}\label{eqn:running_avg_R}
\bar{R}_k[t+1] = \left(1 - \frac{1}{\epsilon}\right)\bar{R}_k[t] + \frac{1}{\epsilon} R_k[t],\quad k \in \bar{\mathcal{K}},
\end{equation}
where $\epsilon \geq 1$ determines the memory length in the averaging process. Here, ${R}_k[t]$ denotes the instantaneous spectral efficiency of UE $k$ at time $t$, extending the definition provided in Section~\ref{subsec:Problem Formulation}.

The PF scheduling objective aims to maximize the PF utility, which is defined as the sum of the logarithms of the average spectral efficiencies, i.e.,
\begin{equation}
    U_{\mathrm{PF}}(\mathcal{M}) = \sum_{k \in \mathcal{M}} \log \bar{R}_k.
\end{equation}
Since maximizing $U_{\mathrm{PF}}$ is equivalent to maximizing a weighted sum-rate objective with weights $w_k[t] = 1/\bar{R}_k[t]$, we formulate PF scheduling as a weighted sum-rate maximization problem \cite{Kushner04, Cui19}. By incorporating these weights into the self-nomination framework, UEs can make fairness-aware feedback decisions, ensuring balanced scheduling over time.
% Consequently, the optimization problem at time $t$ becomes
% \begin{subequations}\label{eqn:problem_formulation_time}
% \begin{align}
% \max_{f(\cdot;\mathbf{\Theta})}
% & \quad
% \mathbb{E}\left[\sum_{m \in \mathcal{M}} w_m[t]\;R_m[t]\right],\\
% \text{subject to}
% & \quad
% \mathbb{E}\left[\sum_{k \in \bar{\mathcal{K}}} f\left(\mathbf{h}_k[t];\mathbf{\Theta}\right)\right] \;\le\; N_{\mathrm{FB}},
% \end{align}
% \end{subequations}
% where $R_m[t]$ is the spectral efficiency of the $m$-th scheduled UE at time $t$, and $R_k[t]=0$ for any $k\notin \mathcal{M}.$ 
Note that when $\epsilon = 1$, the weights enforce a round-robin style scheduling, whereas increasing $\epsilon$ broadens the averaging window in $\bar{R}_k[t]$, pushing the algorithm closer to a classic PF regime that emphasizes long-term fairness.

\subsection{Modifications to the Self-Nomination Network}
Unlike the previous DNN designed for sum-rate maximization with uniform weights in Section~\ref{subsec:DNN Architecture and Gradient Approximation}, the PF setting introduces an additional input: the weight $w_k[t]$. Each UE can obtain this weight either through a control signal from the BS or by performing a local computation. 
% The latter is feasible because, under ZF precoding, inter-user interference is largely suppressed, allowing each UE to estimate its effective rate and derive a corresponding weight with minimal overhead.
% The latter is feasible because, under ZF precoding, inter-user interference is largely suppressed, enabling each UE to locally estimate its effective SINR, which in this case reduces to an SNR, and compute the resulting rate without requiring additional information exchange.

\begin{figure}[t]
\begin{center}
\includegraphics[width=\linewidth]
{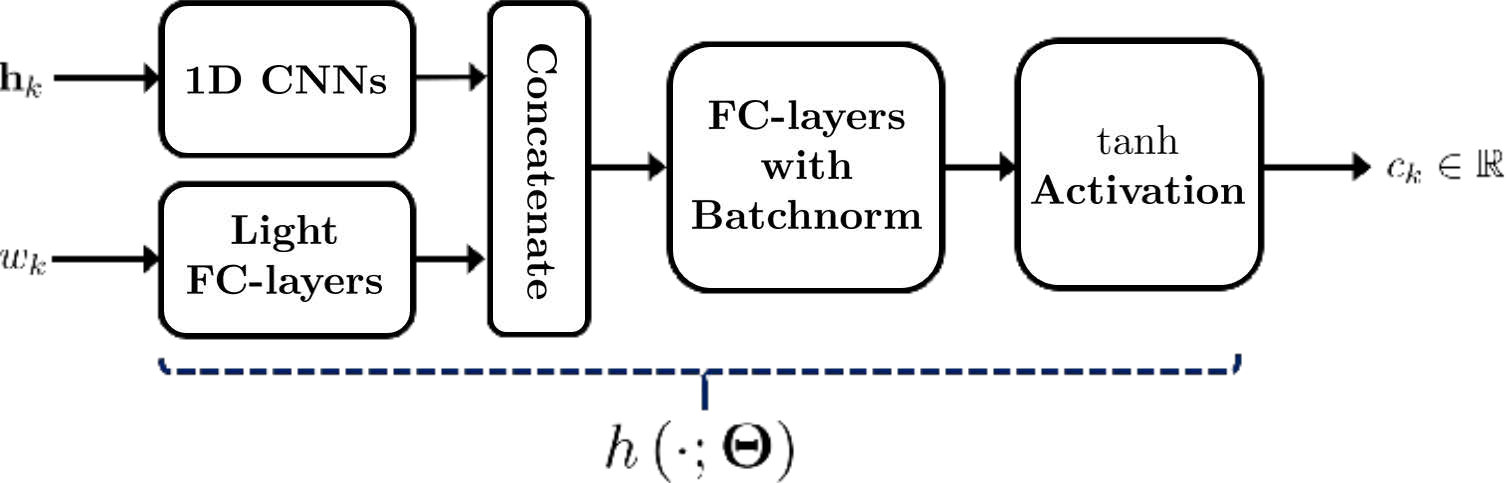}
\end{center}
\caption{The trainable DNN $h(\cdot;\mathbf{\Theta})$ architecture when $\protect\phi(\mathbf{h}_k) = \protect\mathbf{h}_k$ under the PF setup.}
\label{fig:proposed_g_pf}
\end{figure}

Fig.~\ref{fig:proposed_g_pf} illustrates the modified DNN architecture for PF scheduling. The additional input $w_m[t]$ is first processed by a lightweight FC module, and its output is then concatenated with the 1D convolutional features extracted from $\mathbf{h}_k$. The resulting joint feature vector proceeds through further FC layers with batch normalization and a final $\tanh$ activation, producing the self-nomination decision. This design ensures that both channel and weight information are leveraged to make PF-aware feedback decisions.

\subsection{Learning Strategy for Weighted Sum-Rate Maximization}

As noted in \cite{Cui19}, PF scheduling can be highly sensitive to small variations in the PF weights. A slight change in $w_k[t]$ can drastically alter the resulting schedule, making it difficult to train a DNN over the large joint space of channels and weights without getting stuck in local optima. Moreover, collecting reliable PF weights for training is nontrivial. One might attempt to run PF scheduling under an all-feedback scenario, but this would generate data that does not fully capture the partial-feedback setting or the ``unseen" weight realizations arising from less-than-ideal decisions that the DNN also needs to learn from.

To address these issues, we propose a simple random sampling method for generating PF weights during DNN training. Specifically, for each user $k$, we draw $w_k$ independently and uniformly from $[0,1]$. 
% This approach captures a broad range of weight values, providing diverse training samples for the self-nomination DNN. 
This choice is practically justified because the PF weights in a real system, defined by $1/\bar{R}_k[t]$, generally remain within a moderate range. For example, suppose we initialize $\bar{R}_k[1]=1$ for all $k$ and consider a worst-case scenario where a user remains unscheduled, i.e., $R_k[t]$ remains $0$ for $1000$ consecutive time slots under a memory length of $\epsilon=1000$. Then, its weight evolves as $w_k[1000]=\left(1-1/1000\right)^{-1000} \approx 2.72,$
% \begin{equation}
% w_k[1000]=\left(1-\frac{1}{1000}\right)^{-1000} \approx 2.72,
% \end{equation}
which remains reasonably small. This result shows that even in extreme cases where a user remains unscheduled for a long duration, the resulting PF weights stay within moderate values. Thus, the range $[0,1]$ used for training adequately captures the diversity of weights encountered in practice. Although random sampling does not fully replicate every detail of PF scheduling, it provides enough coverage to facilitate effective DNN training under weighted sum-rate maximization, regardless of the specific choice of $\epsilon$.

% Note that $w_k[t]$ in a real PF system is given by $1/\bar{R}_k[t]$, where $\bar{R}_k[t]$ is the running average of user $k$’s spectral efficiency. If we initialize $\bar{R}_k[1]=1$ for all $k$, then $w_k[1]=1$ and gradually decreases for scheduled users while remaining relatively higher for unscheduled ones, preserving the PF principle of eventually serving all users. 
% Although random sampling does not replicate every nuance of PF scheduling, it provides sufficient coverage of weight scenarios to help the DNN generalize effectively under weighted sum-rate maximization.

\section{Performance Analysis} \label{sec:performance_analysis}
We now assess the proposed self-nomination methods for feedback reduction at the UE side. First, we describe the dataset specifications and implementation details. Next, we compare the proposed approach against baseline methods, focusing on downlink sum-rate maximization while varying both the total number of UEs and the available feedback capacity. We also examine the average number of UEs that provide feedback under the proposed methods. Finally, we evaluate the performance of PF scheduling to highlight the applicability of our approach in fairness-oriented scenarios. Throughout this section, we train separate DNNs for each specific setup, including array type, scheduling limit $M$, feedback capacity $N_\mathrm{FB}$, and total number of UEs $\lvert\bar{\mathcal{K}}\rvert$.

\subsection{Dataset Specifications}

The simulation datasets are generated using the QuaDRiGa simulator \cite{Jae23}, beginning with 100{,}000 UE channel realizations derived from the 3GPP 38.901 urban-microcell (UMi) model \cite{ETSI20}. From these realizations, 70,000 channel sets are generated, where each set contains a predetermined number of $\lvert \bar{\mathcal{K}} \rvert$ UE channels randomly selected from the entire pool. Of these sets, 60,000 are used for training and 10,000 for testing. The carrier frequency is set to 7 GHz, representing the upper midband in FR3. Key simulation parameters, such as cell size, base station geometry, channel model, and antenna configurations (ULA and UPA), are summarized in Table~\ref{table:specifications}.

\begin{table}[t]
\centering
\caption{Dataset Specifications}
\label{table:specifications}
\begin{tabular}{|c|c|c|}
\hline
\textbf{Parameter} \rule{0pt}{2.0ex} & \multicolumn{2}{c|}{\textbf{Value}} \\ 
\hline
\hline
Cell type & \multicolumn{2}{c|}{Single cell} \\
\hline
Cell radius & \multicolumn{2}{c|}{100 m} \\
\hline
BS position & \multicolumn{2}{c|}{(0, 0, 10) m} \\
\hline
Channel model & \multicolumn{2}{c|}{UMi in TR 38.901 \cite{ETSI20}} \\
\hline
Carrier frequency & \multicolumn{2}{c|}{7 GHz} \\
\hline
Simulation Bandwidth & \multicolumn{2}{c|}{Narrowband} \\
\hline
\multirow{3}{*}{\vspace{2.5mm} \shortstack{ BS Antenna Setup \vspace{-1mm}\\ (Row $\times$ Column)}} & ULA & 1 × 32 \\
\cline{2-3}
& UPA & 4 × 8 \\
\hline
\end{tabular}
\end{table}

% \subsection{Implementation Details}

\subsection{Downlink Sum-rate Maximization}

% Figure: Sum-Rate vs Baselines
\begin{figure}[t]
\centering
\subfigure[UPA array]{
    \includegraphics[width=\linewidth]
    {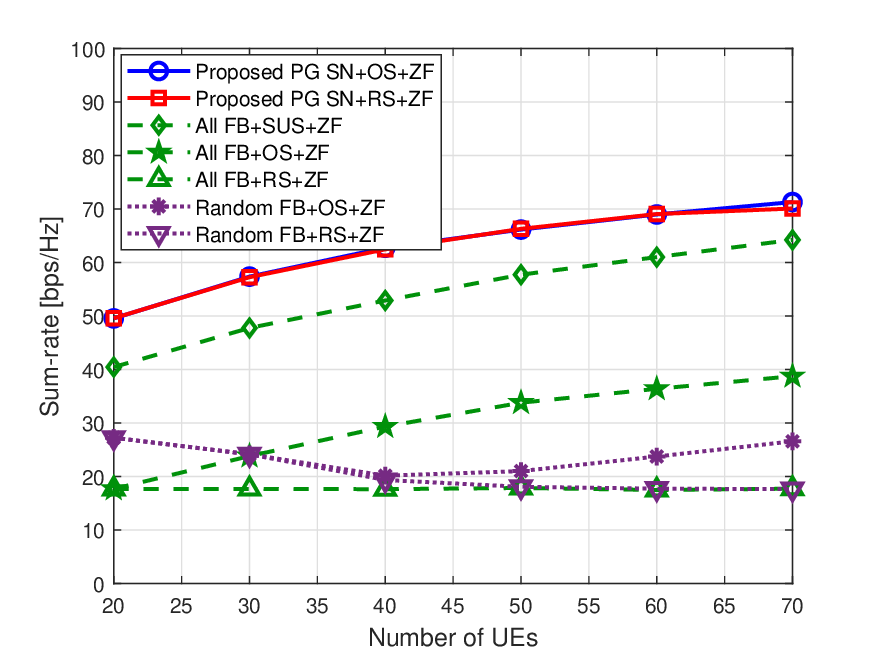}
    \label{fig:UPA_Nfb30_M20_sumrate_vs_baselines}}
\subfigure[ULA array]{
    \includegraphics[width=\linewidth]
    {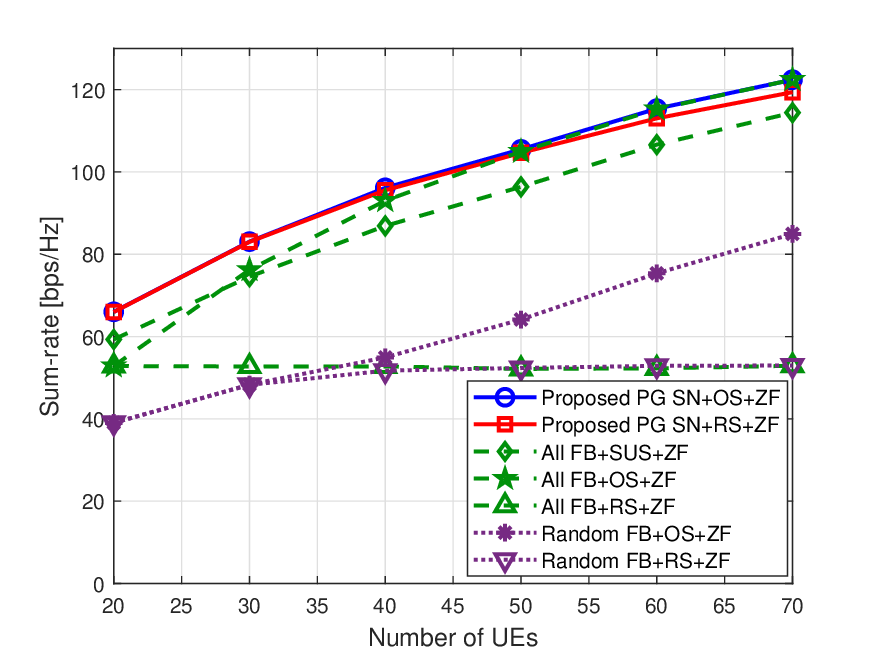}
    \label{fig:ULA_Nfb30_M20_sumrate_vs_baselines}}
\caption{Downlink MU-MIMO sum-rate versus the total number of UEs, evaluated under $N_{\mathrm{FB}}=30$, $M=20$, under two antenna configurations (UPA and ULA).}
\label{fig:sumrate_baselines}
\end{figure}

% Figure: Number of Self-Nominating Users
\begin{figure}[t]
\centering
\subfigure[UPA array]{
    \includegraphics[width=\linewidth]
    {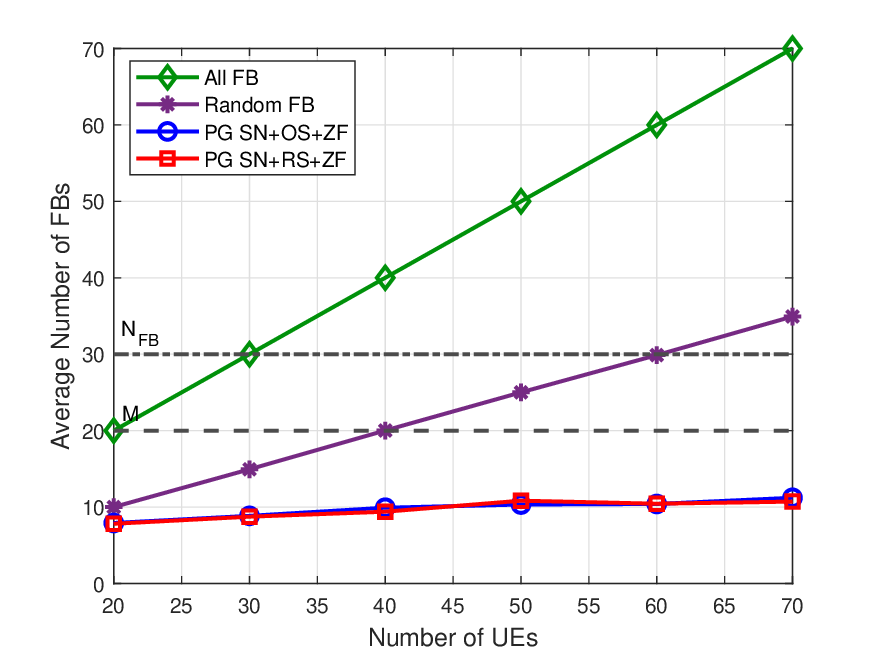}
    \label{fig:UPA_M30_K20_num_SN_users}}
\subfigure[ULA array]{
    \includegraphics[width=\linewidth]
    {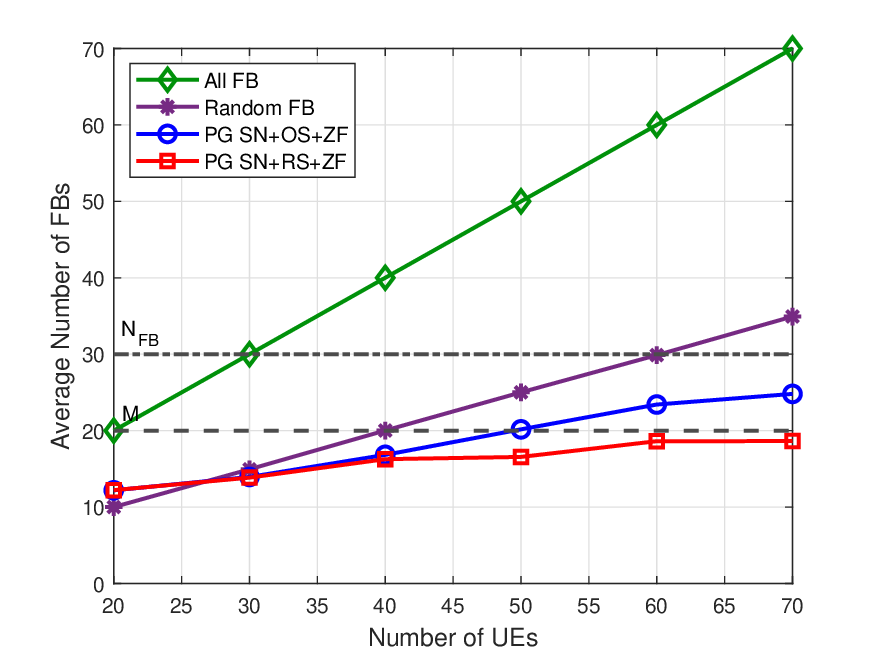}
    \label{fig:ULA_Nfb30_M20_num_SN_users}}
\caption{Average number of self-nominated users versus the total number of UEs, evaluated at $N_{\mathrm{FB}}=30$, $M=20$, und UPA and ULA.}
\label{fig:SN_users_comparison}
\end{figure}

In these simulations, each evaluated approach comprises three components: a UE-side feedback strategy, a BS-side UE scheduling method, and ZF precoding. For feedback, we consider two proposed self-nomination (SN) strategies--direct optimization-based (DO) and policy gradient-based (PG)--along with two baselines: ``All FB," where every user feeds back its CSI, and ``Random FB," where each user randomly decides, with probability 0.5, whether to feed back.
For the baseline methods, we do not constrain the average sum feedback count to $N_\mathrm{FB}$ unless otherwise stated, as the goal is to compare them under the full potential of their feedback strategies. Next, for scheduling, random scheduling (RS), opportunistic scheduling (OS), and semi-orthogonal user selection (SUS) method \cite{Yoo06} are employed. The SUS algorithm iteratively selects UEs whose channels are strong and semi-orthogonal.
Finally, ZF precoding is applied to the chosen UEs. Thus, a label such as ``All FB + OS + ZF" indicates a combination where all users feed back, the BS selects UEs via OS, and precoding is performed by ZF. Likewise, ``Proposed PG SN + RS + ZF" denotes a policy gradient-based self-nomination procedure, random scheduling, and ZF precoding. For enhanced clarity and readability, each paragraph begins with a highlighted key takeaway summarizing the main insight of the corresponding numerical results.

\textbf{The proposed self-nomination methods significantly outperform baselines in the UPA configuration, due to better management of spatial correlation.}
Fig.~\ref{fig:sumrate_baselines} shows the downlink MU-MIMO sum-rate as the total number of UEs varies. Each proposed method is trained specifically with its paired scheduling method. In the UPA configuration (Fig.~\ref{fig:UPA_Nfb30_M20_sumrate_vs_baselines}), both proposed self-nomination methods outperform ``All FB" and ``Random FB" across the entire UE range. This advantage stems from the limited azimuth resolution of UPA channels for the considered array configuration with few horizontal antennas, where carefully selecting spatially compatible UEs becomes more important than simply increasing the number of UE feedbacks.
% In particular, greedily choosing UEs with the strongest channel gains can lead to highly correlated channels and poor ZF performance, making such a strategy suboptimal in this setting. 
Consequently, our self-nomination approach outperforms baseline methods, aligning with the observation that SUS (i.e., ``All FB+SUS+ZF") is also effective in this scenario. 
On the other hand, ``All FB+OS+ZF" performs poorly because it tends to greedily select UEs with the strongest channel gains, which often results in scheduling highly correlated users. This severely degrades ZF precoding performance, resulting in reduced beamforming gain.
Random FB performs relatively well when the number of UEs is small, for a similar reason.

\textbf{The proposed self-nomination methods remain competitive with baseline methods in the ULA setup, even without sophisticated scheduling.}
In contrast, the ULA setup (Fig.~\ref{fig:ULA_Nfb30_M20_sumrate_vs_baselines}) offers higher azimuth resolution, allowing OS to become increasingly effective as the number of UEs grows.
% Nevertheless, the proposed self-nomination remains competitive, as evidenced by ``Proposed PG SN+RS+ZF" performing comparably to both ``Proposed PG SN+OS+ZF" and ``All FB+OS+ZF." Notably, even with a simple random scheduling strategy, strong downlink sum-rate performance can be achieved by controlling UE-side feedback.
Nevertheless, the proposed self-nomination approach remains competitive. Specifically, ``Proposed PG SN+OS+ZF" performs on par with ``All FB+OS+ZF," while ``Proposed PG SN+RS+ZF" also achieves comparable sum-rate performance despite relying on random scheduling. This highlights that strong downlink sum-rate performance can still be achieved through effective control of UE-side feedback, even without sophisticated scheduling.

% Figure: Sum-Rate vs Proposed Methods
\begin{figure}[t]
\centering
\subfigure[UPA array]{
    \includegraphics[width=\linewidth]
    {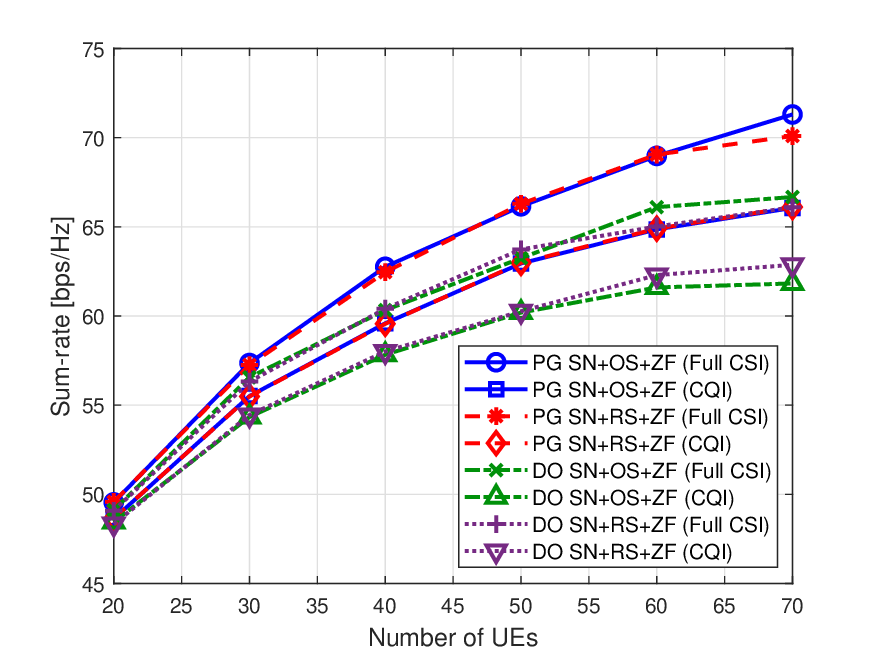}
    \label{fig:UPA_Nfb30_M20_sumrate_vs_proposed_methods}}
\subfigure[ULA array]{
    \includegraphics[width=\linewidth]
    {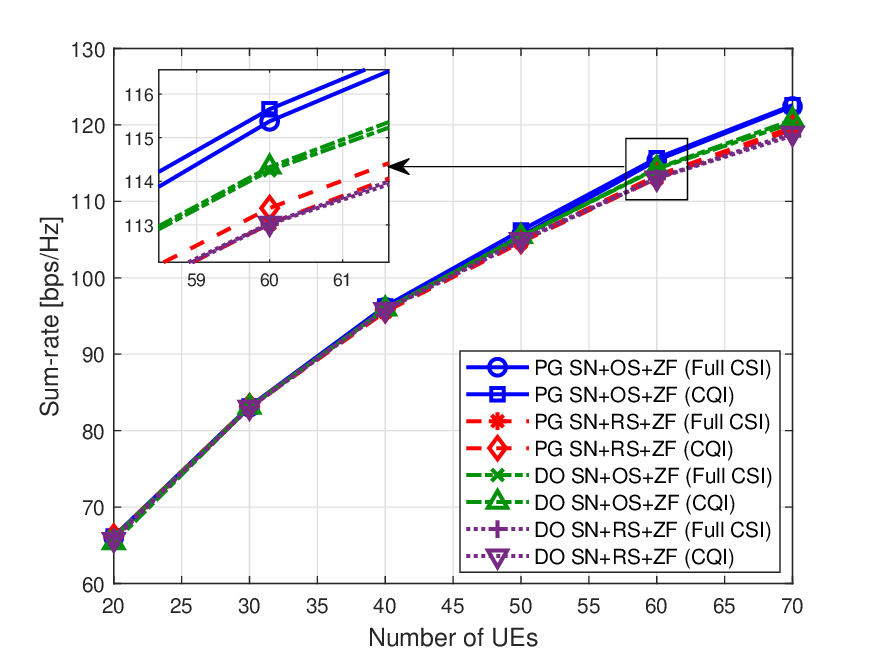}
    \label{fig:ULA_Nfb30_M20_sumrate_vs_proposed_methods}}
\caption{Downlink MU-MIMO sum-rate versus the total number of UEs for the proposed methods, evaluated at $N_{\mathrm{FB}}=30$, $M=20$, under UPA and ULA.}
\label{fig:sumrate_proposed}
\end{figure}

\textbf{Lower azimuth resolution (UPA) leads to fewer self-nominations due to increased spatial congestion, while higher azimuth resolution (ULA) allows more aggressive self-nomination.}
The impact of azimuth resolution on spatial congestion is further illustrated in Fig.~\ref{fig:SN_users_comparison}. In the UPA scenario (Fig.~\ref{fig:UPA_M30_K20_num_SN_users}), the lower azimuth resolution leads to increased spatial congestion, resulting in more correlated channels. As a result, the number of self-nominating UEs remains relatively low, staying around 10, even though more users could be scheduled. In contrast, the ULA scenario (Fig.~\ref{fig:ULA_Nfb30_M20_num_SN_users}) exhibits less congestion and allows for more aggressive feedback, with self-nomination counts increasing with the total number of UEs. Note that PG with OS nominates slightly more than the scheduling limit ($M$), while PG with RS remains below it to avoid poor-channel UEs. This difference in nomination behavior helps explain the performance gap observed in Fig.\ref{fig:ULA_Nfb30_M20_sumrate_vs_baselines}.

\textbf{The proposed self-nomination significantly reduces feedback overhead.}
Despite these differences, both setups demonstrate the feedback efficiency of the proposed method. In the UPA case, self-nomination uses only about $14\%$ of the total $70$ UEs for feedback, while still outperforming the baselines. In the ULA case, even with just $35\%$ of the $70$ UEs feeding back, representing a $65\%$ reduction, ``Proposed PG SN+OS+ZF" matches the sum-rate of ``All FB+OS+ZF," showing that effective self-nomination can achieve competitive performance with substantially reduced overhead.

\textbf{PG-based self-nomination consistently outperforms DO-based methods under UPA, and using full CSI \mbox{inputs} substantially improves sum-rate over CQI-only inputs.}
Fig.~\ref{fig:sumrate_proposed} illustrates the downlink sum-rate performance of the proposed self-nomination strategies, comparing DO versus PG, full CSI input versus CQI-only input, and training under OS versus RS. In the UPA configuration (Fig.~\ref{fig:UPA_Nfb30_M20_sumrate_vs_proposed_methods}), PG-based self-nomination outperforms DO-based self-nomination overall. Furthermore, models using full CSI tend to achieve higher sum-rates than those relying solely on CQI. Meanwhile, the scheduling choice (OS or RS) has a limited impact in the UPA case, since the high spatial correlation naturally limits the number of users that can be effectively scheduled, reducing the influence of the scheduler.

\textbf{Performance differences between input types (full CSI versus CQI), self-nomination methods (PG versus DO), and schedulers (OS versus RS) are minor in the less correlated ULA scenario.}
On the contrary, Fig.~\ref{fig:ULA_Nfb30_M20_sumrate_vs_proposed_methods} shows that when channels are less correlated (as in the ULA configuration), the overall performance differences between full CSI and CQI inputs, PG and DO self-nomination, and OS and RS scheduling remain small. Interestingly, CQI-based methods consistently outperform their full CSI counterparts by around 0.5 bps/Hz when the number of UEs is large (e.g., 60 or 70), suggesting that in scenarios where spatial channel characteristics are less critical, CQI inputs may offer a slight advantage due to their simpler structure.

\begin{figure}[t]
\begin{center}
\includegraphics[width=\linewidth]
{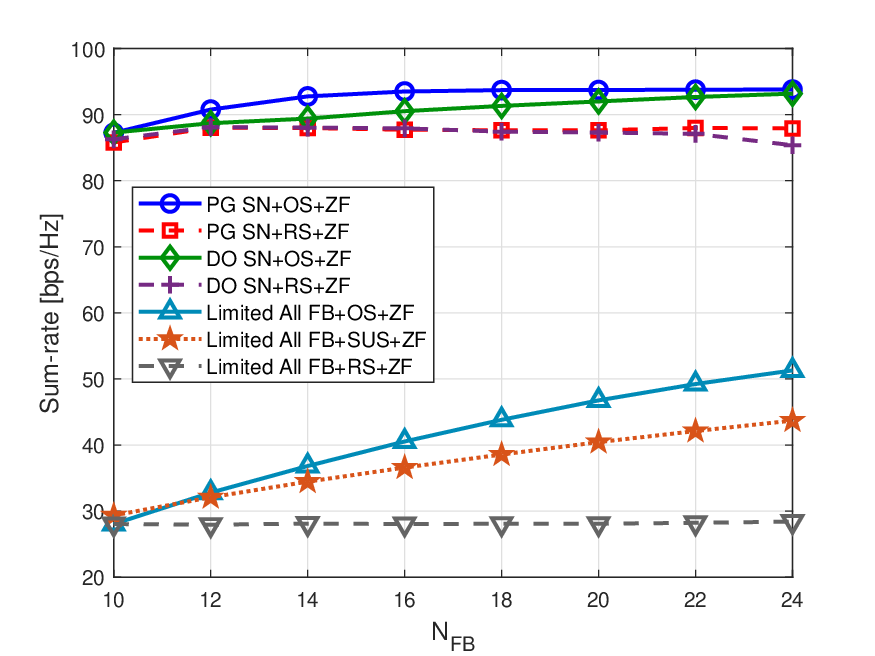}
\end{center}
\caption{Downlink MU-MIMO sum-rate versus $N_{\mathrm{FB}}$, with fixed $M=10$ and $\lvert\bar{\mathcal{K}}\rvert=60$ under the ULA configuration.}
\label{fig:ULA_M10_dl_sumrate_over_Nfb}
\end{figure}

\begin{figure}[t]
\begin{center}
\includegraphics[width=\linewidth]
{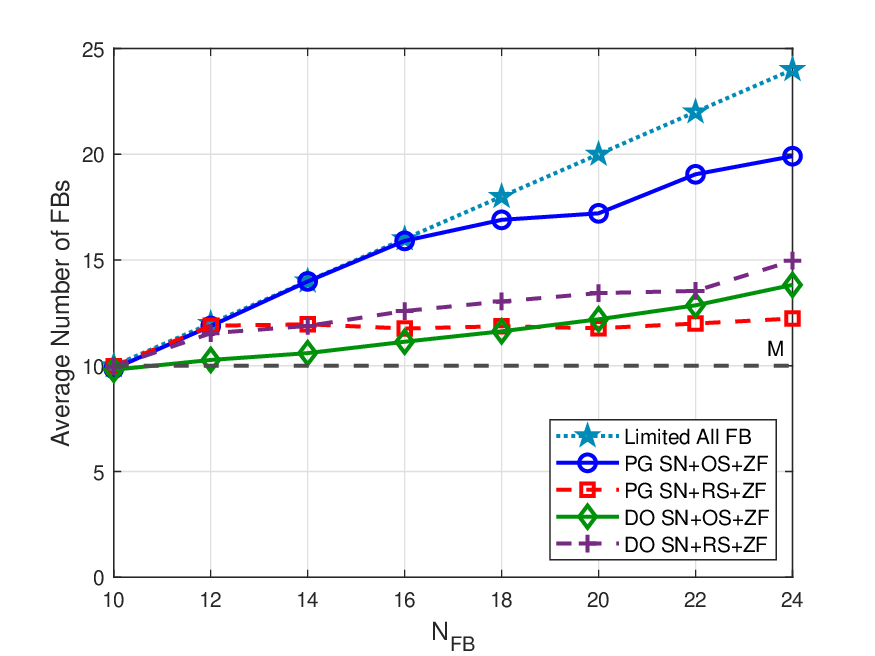}
\end{center}
\caption{Average number of self-nominated users versus $N_{\mathrm{FB}}$, with fixed $M=10$ and $\lvert\bar{\mathcal{K}}\rvert=60$ under the ULA configuration.}
\label{fig:ULA_M10_num_SN_users_over_Nfb}
\end{figure}

\textbf{Without self-nomination, constrained feedback significantly degrades sum-rate performance.}
Fig.~\ref{fig:ULA_M10_dl_sumrate_over_Nfb} presents the downlink sum-rate as a function of the feedback capacity $N_\mathrm{FB}$ under a ULA configuration. The scheduling limit is fixed at $M=10$, meaning up to 10 users are selected from the self-nominated set $\mathcal{K}$, which is drawn from a total pool of $\lvert\bar{\mathcal{K}}\rvert = 60$ UEs. We newly introduce the ``Limited All FB" baseline, where all UEs attempt to feed back their CSI, but if the total exceeds $N_\mathrm{FB}$, a random subset of $N_\mathrm{FB}$ UEs is selected to proceed. This round-robin-like approach leads to substantial sum-rate degradation, highlighting the importance of carefully designed selective feedback mechanisms. Meanwhile, among the proposed methods, ``PG SN+OS+ZF" achieves the highest sum-rate, increasing up to $N_\mathrm{FB}=16$, while ``PG SN+RS+ZF" saturates earlier at $N_\mathrm{FB}=14$. ``DO SN+OS+ZF" improves more gradually across the entire feedback range. 
These trends are discussed next by examining how the average number of feedback UEs varies with $N_\mathrm{FB}$ in Fig.~\ref{fig:ULA_M10_num_SN_users_over_Nfb}.

% ``DO SN+RS+ZF" initially follows a similar trend to its PG counterpart but exhibits a slight drop at higher $N_\mathrm{FB}$ values, potentially due to less effective feedback-user matching under random scheduling.}

\textbf{PG-based SN combined with OS better exploits increased feedback capacity compared to the overly conservative DO-based methods.}
These sum-rate trends are further explained by Fig.~\ref{fig:ULA_M10_num_SN_users_over_Nfb}, which shows the average number of UEs that provide feedback as $N_\mathrm{FB}$ varies. Specifically, ``PG SN+OS+ZF" consistently uses the full feedback budget up to $N_\mathrm{FB}=16$, effectively leveraging additional feedback opportunities.
Similarly, ``PG SN+RS+ZF" achieves its best performance when approximately 12 UEs feed back, quickly adjusting as the feedback constraint relaxes between $N_\mathrm{FB}=10$ and $12$.
On the other hand, both ``DO SN+RS+ZF" and ``DO SN+OS+ZF" respond conservatively as $N_\mathrm{FB}$ increases, failing to fully utilize the available feedback. This conservative behavior suggests that the direct optimization approach does not closely track the optimal solution as the feedback constraint is relaxed.
% become less stringent.
% The slight performance drop of ``DO SN+RS+ZF" near $N_\mathrm{FB}=24$ (see Fig.~\ref{fig: ULA_M10_dl_sumrate_over_Nfb}) suggests that DO-based methods adapt more slowly to changes in $N_\mathrm{FB}$ compared to PG-based approaches.

\begin{figure}[t]
\begin{center}
\includegraphics[width=\linewidth]
{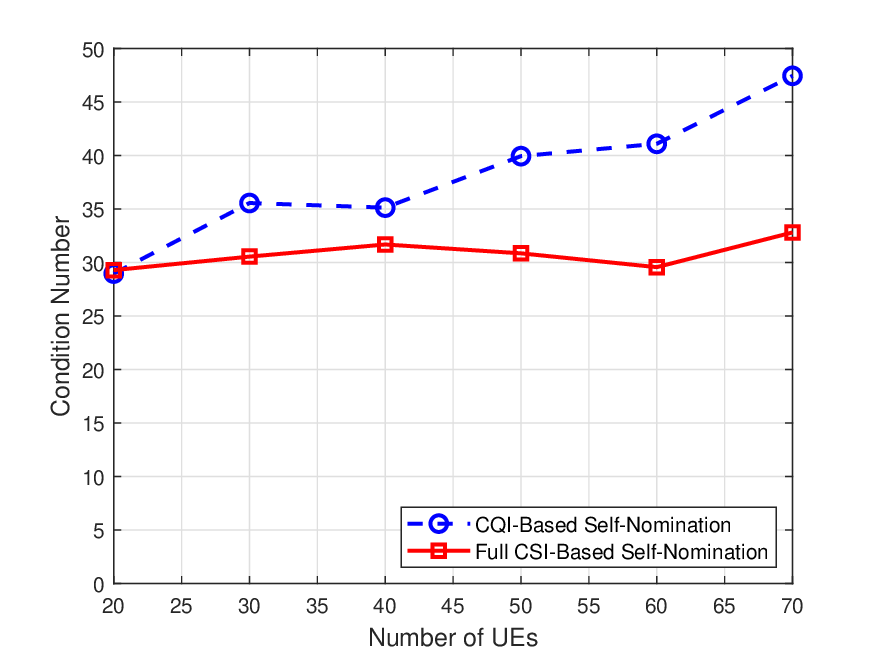}
\end{center}
\caption{Comparison of the condition number for the self-nominated set $\mathcal{K}$ under CQI-based and full CSI-based policies, with $M = 20$ and $N_\mathrm{FB} = 30$ in the UPA configuration.}
\label{fig:condition_number}
\end{figure}

\textbf{Full CSI-based self-nomination captures both spatial characteristics and channel strength, enabling better spatial compatibility compared to CQI-based self-nomination.}
Fig.\ref{fig:condition_number} compares the condition numbers of the PG-based self-nomination method using either CQI or full CSI as input, assuming $M = 20$ and $N_\mathrm{FB} = 30$ under the UPA setup, where spatial congestion is significant.
The condition number, derived from the matrix $\widetilde{\mathbf{H}} \triangleq \bigl[\mathbf{h}_1,\ldots,\mathbf{h}_{\lvert \mathcal{K} \rvert}\bigr]^H$ based on the self-nominated UE set $\mathcal{K}$, serves as a metric for evaluating both the spatial compatibility and overall magnitude of the selected downlink channels \cite[Section V-C]{Castaneda17}.
Results show that CQI-based self-nomination yields higher and increasing condition numbers as the number of UEs grows, indicating greater spatial correlation among selected users. 
% Despite achieving a much lower condition number than the centralized ``All FB" baseline, which reaches around 855 at $\lvert\bar{\mathcal{K}}\rvert = 60$, the CQI-based approach remains less effective than full CSI. To be specific, 
Full CSI-based self-nomination consistently maintains a low and stable condition number around 30, demonstrating its ability to leverage detailed spatial information, which the CQI-based approach lacks. This leads to more effective user selection and improved beamforming performance.

\subsection{Proportional-Fair Scheduling}
Fairness is evaluated by considering 10 wireless network layouts over either 100 or 1000 time slots, using two different values of $\epsilon$ (100 and 1000). Recall that only the policy gradient-based method is used in this setup, so any reference to the ``proposed self-nomination method" here denotes the policy gradient-based approach. A block fading model is assumed, whereby channel states vary across time slots according to the 3GPP UMi distribution with a ULA. The mean rate of each link (UE) is defined as the average of its instantaneous rates across time slots, which can be expressed as
\begin{equation}
\widetilde{R}_k = \frac{1}{T}\sum_{t=1}^T R_k[t].
\end{equation}
Note that this time-averaged $\widetilde{R}_k$ differs from the running average $\bar{R}_k$ used to derive PF weights in \eqref{eqn:running_avg_R}.

\begin{figure}[t]
\begin{center}
\includegraphics[width=\linewidth]
{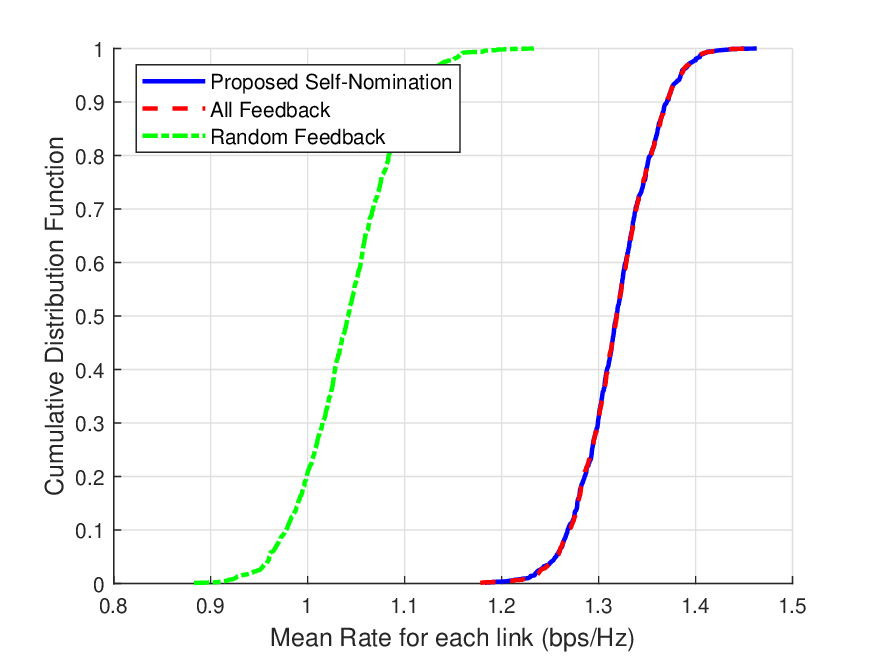}
\end{center}
\caption{CDF of mean rates for 100 UEs, with $N_{\mathrm{FB}}=40$ and $M=20$, using a ULA.}
\label{fig:ULA_PF_CDF_Nfb40_T1000}
\end{figure}

\begin{table}[t]
    \centering
    \caption{Mean Log Utility and Feedback Count Comparison for Different Memory Lengths ($\epsilon$) and Time Steps ($T$)}
    \label{tab:pf_results}
    \begin{tabular}{cc|ccc|ccc}
        \toprule
        \multirow{2}{*}{$\epsilon$} & \multirow{2}{*}{$T$} & \multicolumn{3}{c|}{\textbf{Log Utility}} & \multicolumn{3}{c}{\textbf{Feedback Count}} \\
        % \cmidrule(lr){3-7}
        & & SN & Random & All & SN & Random & All \\
        \midrule
        10  & 100  & 22.48  & -1.33 & 21.76  & 36.36  & 49.77 & 100.00\\
        10  & 1000   & 25.32  & 0.23  & 24.42  & 36.24  & 49.92 & 100.00 \\
        100 & 100  & 12.09  & 1.16  & 12.12  & 36.37  & 49.77 & 100.00 \\
        100 & 1000  & 27.65  & 3.99  & 27.67  & 35.88  & 49.92 & 100.00 \\
        \bottomrule
    \end{tabular}
\end{table}

Fig.~\ref{fig:ULA_PF_CDF_Nfb40_T1000} compares the cumulative distribution function (CDF) of each link’s mean rate under three methods: the proposed self-nomination, random feedback at probability 0.5, and all feedback, with $\epsilon=100$ and $T=1000$. All three methods exhibit a similar slope and trend because PF scheduling and ZF precoding are applied after feedback. However, their CDFs show a clear horizontal shift. Notably, self-nomination achieves a mean-rate distribution comparable to all feedback but with significantly fewer transmissions. In contrast, random feedback results in a noticeable performance gap relative to both self-nomination and all feedback, highlighting that reducing UE-side feedback arbitrarily degrades mean rate performance. These results confirm that self-nomination effectively reduces feedback overhead while maintaining performance close to the full-feedback scenario.

Table~\ref{tab:pf_results} summarizes the mean log utility and average feedback counts achieved by the proposed self-nomination (SN) method, random feedback, and all-feedback scenarios for varying memory lengths ($\epsilon$) and simulation durations ($T$). The mean log utility of self-nomination closely aligns with that of the full-feedback baseline across all evaluated cases but requires significantly fewer CSI transmissions. For example, when $\epsilon=100$ and $T=1000$, self-nomination achieves a mean log utility of 27.65, nearly identical to the $27.67$ obtained by all-feedback, yet reduces the average feedback count by approximately $65\%$. In contrast, the random feedback strategy substantially degrades mean log utility, highlighting the importance of intelligent feedback decisions. Notably, self-nomination consistently outperforms the random feedback method across all considered scenarios, regardless of the memory length ($\epsilon$) or duration ($T$). Specifically, smaller $\epsilon$ values reflect shorter memory windows, leading to quicker fairness convergence, whereas larger $\epsilon$ values imply longer memory windows, thus requiring more time steps for fairness stabilization. Despite these diverse conditions, the self-nomination network trained solely through random weight sampling effectively generalized, demonstrating its robustness and superior adaptability to varying PF scheduling scenarios. These results clearly illustrate the effectiveness of self-nomination in preserving fairness and efficiently managing feedback overhead under realistic and dynamic conditions.

\section{Conclusion} \label{sec:conclusion}
This paper presented a deep learning-based self-nomination framework to mitigate CSI feedback overhead in MU-MIMO systems. The results demonstrated that allowing UEs to independently decide their feedback significantly reduced redundant CSI transmissions without degrading sum-rate performance. Compared to CQI-based thresholding, self-nomination with full CSI leveraged spatial channel information more effectively, leading to improved user selection and scheduling. The proposed optimization techniques---direct Lagrangian and policy gradient-based approaches---both successfully trained self-nomination networks under average feedback constraints, with the policy gradient method exhibiting greater adaptability to variations in feedback capacity. Furthermore, by extending self-nomination to proportional-fair scheduling, we showed that fairness-aware feedback decisions can also be effectively incorporated.

A key additional advantage of self-nomination is that UEs not selected to feed back can simply remain silent, which provides the potential for significant power reduction at the UE side.   Although we have not attempted to quantify the power reduction, completely silencing a UE not only yields transmit power savings but also provides an opportunity to enter sleep modes, provided no data transmission is pending.  Finally, we note that self-nomination is complementary to deep learning-based CSI compression, which is under active standardization in 3GPP. It can be directly integrated with these compression techniques to further reduce feedback overhead and energy consumption, making it a practical solution for future wireless networks.

\begin{appendices}
\section{Proof of Proposition~\ref{prop:policy_gradient}}
\label{app:proof_prop_policy_gradient}
Let $\mathcal{A} = \{0,1\}^{|\bar{\mathcal{K}}|}$ denote the action space, consisting of all possible binary feedback decisions. We begin with the definition of the gradient of the expected Lagrangian:
\begin{equation*} 
\nabla_{\mathbf{\Theta}} \mathbb{E}_{\pi} \left[ \mathcal{L}(\{\mathbf{h}_k\},\mathbf{a}) \right] = \nabla_{\mathbf{\Theta}} \sum_{\mathbf{a} \in \mathcal{A}} \pi(\mathbf{a} \mid \{\mathbf{h}_k\}; \mathbf{\Theta}) \mathcal{L}(\{\mathbf{h}_k\}, \mathbf{a}). 
\end{equation*} 
Since summation and differentiation are interchangeable under mild regularity conditions, we rewrite the gradient as
\begin{equation*} 
\sum_{\mathbf{a} \in \mathcal{A}} \nabla_{\mathbf{\Theta}} \pi(\mathbf{a} \mid \{\mathbf{h}_k\}; \mathbf{\Theta}) \cdot \mathcal{L}(\{\mathbf{h}_k\}, \mathbf{a}). 
\end{equation*}

To proceed, we apply the log-derivative trick, which states that for any differentiable probability distribution $\pi$,
\begin{equation*} 
\nabla_{\mathbf{\Theta}} \pi(\mathbf{a} \mid \{\mathbf{h}_k\}; \mathbf{\Theta}) = \pi(\mathbf{a} \mid \{\mathbf{h}_k\}; \mathbf{\Theta})  \cdot \nabla_{\mathbf{\Theta}} \log \pi(\mathbf{a} \mid \{\mathbf{h}_k\}; \mathbf{\Theta}). 
\end{equation*}
Substituting this into the summation, we obtain
\begin{equation*} 
\sum_{\mathbf{a} \in \mathcal{A}} \pi(\mathbf{a} \mid \{\mathbf{h}_k\}; \mathbf{\Theta})  \cdot \nabla_{\mathbf{\Theta}} \log \pi(\mathbf{a} \mid \{\mathbf{h}_k\}; \mathbf{\Theta})  \cdot \mathcal{L}(\{\mathbf{h}_k\}, \mathbf{a}). 
\end{equation*}
Recognizing that the summation represents the expectation with respect to the policy distribution $\pi$, we conclude that
\begin{equation*} 
\nabla_{\mathbf{\Theta}} \mathbb{E}_{\pi} \left[ \mathcal{L}(\{\mathbf{h}_k\},\mathbf{\Theta}) \right] \!=\! \mathbb{E}_{\pi} \left[ \nabla_{\mathbf{\Theta}} \log \pi(\mathbf{a} \mid \{\mathbf{h}_k\}; \mathbf{\Theta}) \!\cdot\! \mathcal{L}(\{\mathbf{h}_k\}, \mathbf{\Theta}) \right]. 
\end{equation*}
This completes the proof.

\end{appendices}

% \section*{Acknowledgment}
% The authors would like to thank Jinfeng Du and Harish Viswanathan for their helpful suggestions, discussions, and feedback, and Jie Chen for providing high-quality upper midband channel datasets.

\bibliographystyle{ieeetr}
\begingroup
\bibliography{JuseongREF}
\endgroup

\end{document}